\documentclass[12pt]{article}

\pdfoutput=1
\usepackage[latin9]{inputenc}
\usepackage[top=80pt,bottom=85pt,left=67pt,right=67pt]{geometry}
\usepackage{amssymb,amsmath}
\usepackage{xypic,subfigure}
\usepackage{graphicx,color,xcolor}
\usepackage{bbm}
\usepackage{float}
\usepackage{cite}
\usepackage{slashed}
\usepackage{breqn}
\catcode`_=12
\catcode`^=12
\usepackage{tensor}
\usepackage[debug,pageanchor=false]{hyperref}
\definecolor{link}{rgb}{.8,.15,.1}
\hypersetup{colorlinks=true,linkcolor=link,citecolor=link,urlcolor=link,linktocpage}

\usepackage{tikz-cd}

 \allowdisplaybreaks

\makeatletter
\@addtoreset{equation}{section}
\makeatother

\newcommand{\beq}{\begin{equation}}
\newcommand{\eeq}{\end{equation}}
\newcommand{\bea}{\begin{eqnarray}}
\newcommand{\eea}{\end{eqnarray}}
\newcommand{\nn}{\nonumber}

\newcommand{\eq}{\begin{equation}}
\newcommand{\feq}{\end{equation}}
\newcommand{\eqn}{\begin{eqnarray}}
\newcommand{\feqn}{\end{eqnarray}}

\usepackage[bbgreekl]{mathbbol}
\usepackage{bbold}
\DeclareSymbolFontAlphabet{\mathbb}{AMSb}
\DeclareSymbolFontAlphabet{\mathbbl}{bbold}
\newcommand{\spindle}{\mathbbl{\Sigma}}

\begin{document}
\begin{titlepage}

\begin{center}

\vskip .5in 
\noindent

{\Large \bf{Monodromy Defects in Massive Type IIA}}

\bigskip\medskip

Andrea Conti$^{a}$\footnote{contiandrea@uniovi.es}, Yolanda Lozano$^a$\footnote{ylozano@uniovi.es}, Christopher Rosen$^b$\footnote{rosen@physics.uoc.gr} \\

\bigskip\medskip
{\small 

a: Department of Physics, University of Oviedo,
Avda. Calvo Sotelo s/n, 33007 Oviedo}

\medskip
{\small and}

\medskip
{\small 

Instituto Universitario de Ciencias y Tecnolog\'ias Espaciales de Asturias (ICTEA),\\
Calle de la Independencia 13, 33004 Oviedo, Spain}

\medskip
{\small 

b: Crete Center for Theoretical Physics,  Institute for Theoretical 
and Computational Physics, \\Department of Physics, 
University of Crete, 71003 Heraklion, Greece}

\vskip 2cm 

     	{\bf Abstract }
	\end{center}
	\noindent

In this paper we study solutions to massive Type IIA supergravity which we propose are dual to co-dimension 2 monodromy defects in 6d (1,0) CFTs realised in NS5-D6-D8 brane systems, as well as in the 5d Sp(N) fixed point theory. In the first case the defects are studied holographically as solutions to 7d $U(1)$ gauged supergravity that asymptote locally to its maximally supersymmetric $\text{AdS}_7$ vacuum away from the defects. In the second case they are dual to solutions to 6d $U(1)^2$ gauged supergravity that asymptote to its maximally supersymmetric $\text{AdS}_6$ vacuum. These solutions are then uplifted to massive Type IIA supergravity using known consistent truncations previously constructed in the literature. We compute the defect entanglement entropy and provide evidence that for the 3d defects, the entanglement entropy can be written as a linear combination of the free energy and conformal weight of the defect. Finally, we construct new co-dimension 2 monodromy defects in 6d and 5d CFTs compactified on Riemann surfaces and/or spindles.
    	
\noindent

\vfill
\eject

\end{titlepage}

\tableofcontents

\section{Introduction}

The study of defect conformal field theories using holography has become a fruitful line of investigation in string theory.  Defect CFTs are typically realised when some of the conformal isometries of a higher dimensional CFT, the so-called ambient CFT, are broken by the introduction of a position-dependent coupling, such that a subgroup of the conformal isometries remains. If the ambient CFT admits a holographic dual the defects can be studied introducing probe branes realising the unbroken symmetries of the defect inside the AdS$_d$ vacuum dual to the ambient CFT. This description is valid when the number of degrees of freedom associated to the defect is small. However, when this is not the case it is necessary to account for their backreaction in the geometry. The deformations induced by the position-dependent couplings are often realised geometrically by extra branes that contribute to the brane intersection where the ambient CFT lives, giving rise to a lower dimensional brane intersection from which a lower dimensional AdS$_{d-p}$ solution arises upon taking the near horizon limit. This lower dimensional AdS background contains a non-trivial warping between the AdS$_{d-p}$ subspace and the internal manifold, dependent on a non-compact coordinate. When this non-compact coordinate is sent to infinity the higher dimensional AdS$_d$ space dual to the ambient CFT arises, in its parametrisation as a foliation of the lower dimensional AdS$_{d-p}$ space times a $p$-dimensional sphere over the non-compact direction. The presence of this non-compact direction renders the holographic interpretation of the CFT dual to the lower dimensional AdS$_{d-p}$ solution subtle. The presence of these subtleties can be seen as a manifestation of the idea that the defect CFT makes no sense by itself far away from the defects, where it must be completed by the higher dimensional CFT where the defects are embedded.

To date, many AdS$_{d-p}$ backgrounds with non-compact directions in the internal space have been interpreted as describing defect CFTs with different co-dimensionality (see \cite{DHoker:2007zhm,DHoker:2007hhe,DHoker:2007mci,DHoker:2008lup,DHoker:2008rje,Dibitetto:2017tve,Dibitetto:2017klx,Dibitetto:2018iar,Dibitetto:2018gtk,Gutperle:2018fea,Gutperle:2019dqf,Chen:2019qib,Chen:2020mtv,Faedo:2020nol,Lozano:2020sae,Faedo:2020lyw,Gutperle:2020rty,Lozano:2021fkk,Gutperle:2022pgw,Lozano:2022vsv,Lozano:2022swp,Conti:2023naw,Gutperle:2023yrd,Capuozzo:2024onf,Lozano:2024idt,Arav:2024exg,Karndumri:2024jib,Karndumri:2024gtv,Lozano:2022ouq,Conti:2024rwd,Arav:2024wyg,Bomans:2024vii,Faedo:2025kjf})\footnote{Or, in some instances, deconstructed higher dimensional CFTs \cite{Conti:2024qgx}.}. In some cases the underlying brane intersection has been identified, and shown to consist of the two sets of, ambient and defect, branes mentioned above (see \cite{Dibitetto:2017tve,Dibitetto:2017klx,Dibitetto:2018iar,Dibitetto:2018gtk,Faedo:2020nol,Lozano:2020sae,Faedo:2020lyw,Lozano:2021fkk,Lozano:2022vsv,Lozano:2022swp,Conti:2023naw,Lozano:2024idt,Lozano:2022ouq,Conti:2024rwd,Faedo:2025kjf}). The  defect branes end on the bound state formed by the ambient branes, known to be described by a higher dimensional AdS$_d$ vacuum in the near-horizon limit. The defects describe then the boundary conditions associated to the intersection between the defect branes and this bound state.

A fruitful approach to the study of holographic defects is in the context of a lower $d$-dimensional supergravity admitting an $\text{AdS}_d$ as a vacuum solution. For co-dimension $p$ defects one looks for fibrations of $\text{AdS}_{d-p}\times S^{p-1}$ over an interval\footnote{One can also look at $\text{AdS}_{d-p}\times S^{p-2}$ fibrations on a Riemann surface with boundary.}, and imposes that away from the defects the $\text{AdS}_d$ vacuum arises asymptotically, typically deformed by subleading background fluxes associated to the defect branes. Upon uplift to Type II or eleven dimensional supergravity the brane intersection that underlies the defect solution can then be identified, and from it one can proceed to explore the field theory living in it, from which the defect CFT will arise, typically in the IR.

Recently, co-dimension 2 defects have been studied taking as starting points the spindle compactifications constructed in \cite{Ferrero:2020laf,Ferrero:2020twa,Boido:2021szx,Ferrero:2021wvk,Ferrero:2021ovq,Couzens:2021rlk,Faedo:2021nub,Ferrero:2021etw}. In the spindle constructions one has a finite interval  and an $S^1$, that closes off at both ends of the interval, giving rise to a topological 2-sphere with two conical deficits at the north and south poles, the spindle. Taking a different global completion, namely, the interval to be semi-compact, it is possible to interpret the same local solutions as describing, instead, co-dimension 2 defects \cite{Gutperle:2022pgw,Gutperle:2023yrd,Arav:2024exg,Arav:2024wyg,Bomans:2024vii}. The interpretation of these defects is as monodromy defects, where background gauge fields of an abelian subgroup of the global symmetry induce non-trivial monodromies for fields charged under the global symmetry, when those encircle the defect. These defects have received a lot of attention in recent years, see for instance \cite{Gutperle:2023yrd,Gutperle:2018fea,Gutperle:2019dqf,Bianchi:2019sxz,Bianchi:2021snj,Arav:2024exg,Arav:2024wyg,Giombi:2021uae}. Here as in \cite{Arav:2024exg,Arav:2024wyg} we will allow as well for the possibility of a conical singularity at the origin. Indeed, we will see that in some of our constructions this will be necessary to obtain a monodromy defect interpretation\footnote{In particular, this conical singularity is necessary if we want to have a monodromy implemented by the R-symmetry gauge fields, as will become clear later in our analysis.}.
In some of the previous works defects within 3d, 4d and 6d CFTs have been constructed, that can be related to the spindle solutions to 4d $U(1)^4$, 5d $U(1)^3$ and 7d $U(1)^2$ gauged supergravities. In \cite{Arav:2024exg,Arav:2024wyg,Bomans:2024vii} the solutions have known uplifts to Type IIA or Type IIB supergravities and are interpreted as describing monodromy defects within the ABJM theory or 4d $\mathcal{N}=4$ SYM. In \cite{Gutperle:2023yrd} the 7d solution was uplifted to eleven dimensional supergravity and interpreted as describing monodromy defects within the 6d (2,0) CFT. 

In this paper we extend these results with the study of monodromy defects within 6d (1,0) CFTs realised in NS5-D6-D8 brane intersections, and the 5d Sp(N) fixed point theory. Both CFTs are dual to solutions of massive Type IIA supergravity, namely, the $\text{AdS}_7$ solutions constructed in \cite{Apruzzi:2013yva} and the $\text{AdS}_6$ Brandhuber-Oz solution \cite{Brandhuber:1999np}. The massive IIA theory will be our common denominator, so we will restrict our attention to lower dimensional supergravities which admit oxidation to massive Type IIA supergravity. For this reason our study of monodromy defects in 7d is limited to solutions to $U(1)$ gauged supergravity. Furthermore, we extend results in the literature involving compactifications on a spindle and a Riemann surface, or two spindles, by replacing the spindles by semi-compact $S^1\times I$ spaces. 

An important result of this work concerns the study of the defect entanglement entropy, which follows the holographic prescription proposed in \cite{Ryu:2006bv}. It is well-known that this quantity can be written in terms of the defect Weyl anomaly and the ``defect conformal weight'' for 2d and 4d defects \cite{Jensen:2018rxu,Kobayashi:2018lil,Chalabi:2021jud}, and it was recently shown that this is also the case for 1d superconformal monodromy defects in ABJM \cite{Conti:2025wwf}, where the role of the Weyl anomaly is now played by the ``defect free energy''. In this article we provide strong hints that such a relation also holds for 3d superconformal monodromy defects in the 5d Sp(N) fixed point theory, and argue that it could hold more generally along the lines of \cite{Jensen:2018rxu,Kobayashi:2018lil,Chalabi:2021jud}.

The paper is structured as follows. In section \ref{subsec:dEE} we summarise the prescription employed in \cite{Conti:2025wwf} for computing the defect entanglement entropy for co-dimension 2 monodromy defects. In section \ref{6dtheories} we study co-dimension 2 monodromy defects within the 6d (1,0) CFTs dual to the $\text{AdS}_7$ solutions of \cite{Apruzzi:2013yva}. We begin with the study of holographic monodromy defects in 7d $U(1)$ gauged supergravity, which we ultimately uplift to massive Type IIA supergravity. This allows us to identify the underlying brane set-up and to collect useful information for the detailed identification of the dual field theory. Inspired by the construction in \cite{Boido:2021szx} we then present a solution to 7d U(1) gauged supergravity that, once uplifted to massive Type IIA, we can interpret as describing defects within the 4d $\mathcal{N}=1$ CFTs obtained by compactifying the 6d (1,0) CFTs on a hyperbolic plane, the duals of which were constructed in \cite{Apruzzi:2015zna}. We speculate about the possible existence of an RG flow relating these monodromy defects to the ones living in the 6d (1,0) CFTs.

In section \ref{5dtheories} we perform a similar analysis within the 5d Sp(N) fixed point theory. In this case we are able to discuss a richer class of solutions to 6d $U(1)^2$ gauged supergravity, thanks to the fact that its uplift to massive Type IIA supergravity is now known. We interpret our solutions as describing defects within the 5d Sp(N) fixed point theory. We compute the defect entanglement entropy and relate it to a linear combination of a defect ``free energy'' term and the conformal weight of the defect. We extend these constructions to solutions interpreted as monodromy defects within the 5d Sp(N) fixed point theory compactified on a Riemann surface with constant curvature. As we work within 6d $U(1)^2$ gauged supergravity we can take the Riemann surface to be a sphere, a torus or a hyperbolic plane. We complete this analysis with the construction of a solution that we interpret as describing monodromy defects within the 5d Sp(N) fixed point theory compactified on a spindle, after flowing to the IR. It is plausible that RG flows exist relating the monodromy defects within the 5d Sp(N) fixed point theory compactified on a Riemann surface or a spindle and the monodromy defects within the 5d Sp(N) fixed point theory. This section heavily relies on previous results obtained in \cite{Faedo:2021nub} involving spindle and Riemann surface compactifications of the 5d Sp(N) fixed point theory.  

Section \ref{conclusions} contains our conclusions and future directions. Finally, in Appendix \ref{appendix} we present a collection of related results in M-theory. Specifically, we construct solutions describing monodromy defects within compactifications of the 6d (2,0) theory on a Riemann surface with constant curvature or a spindle.

\section{Entanglement entropy for co-dimension 2 monodromy defects}\label{subsec:dEE}

In this section we briefly summarise the prescription for computing the universal part of the entanglement entropy for co-dimension 2 monodromy defects put forward in \cite{Conti:2025wwf}. 

We consider co-dimension 2 monodromy defects characterised by a metric
\begin{equation}\label{eq:gAnz}
d s^2_{d+1} = e^{2V} d s^2\left(\mathrm{AdS}_{d-1} \right) + h^2 d \varphi^2 + f^2 d y^2,
\end{equation}
where $\varphi$ is a periodic coordinate with period $2\pi$, and $V,h$ and $f$ are functions of $y$, which lives in a semi-compact interval $[y_{core},\infty)$, such that when it approaches infinity the maximally supersymmetric $\mathrm{AdS}_{d+1}$ vacuum of the corresponding $(d+1)$-dimensional supergravity arises\footnote{The meaning of $y_{core}$ will become clear shortly.}.  
A prescription for computing the entanglement entropy for a spherical entangling region centered on the defect was proposed in \cite{Jensen:2013lxa}. The expression derived there exhibits a divergence structure which depends on both the dimensionality of the ambient CFT and on the co-dimension of the defects \cite{Jensen:2013lxa,Affleck:1991tk}. These divergences can be regulated by introducing a short distance cut-off $\tilde{\epsilon}$. For the co-dimension 2 defects we study, the universal part of the entanglement entropy (that which is invariant under rescalings of $\tilde{\epsilon}$) is given by\footnote{$\lceil . \rceil$ is the ceiling function.}
\begin{equation}\label{eq:Ceq}
\mathcal{C}_{\mathcal{D}}^{(d)} = -2^{\lceil\frac{d}{3}\rceil}\frac{(-\pi)^{\lceil\frac{d}{4} \rceil}}{4G_N^{(d+1)}}\left[ \int d y\, e^{(d-3)V} | f h |\right]_{\mathcal{D}} \qquad \mathrm{for}\quad 3\le d \le 6.
\end{equation}
This universal quantity appears in the expansion of the defect entanglement entropy in the limit that $\tilde{\epsilon}\to0$, as
\begin{equation}
\Delta \mathcal{S}_{EE} = c_{d-4} \frac{R^{d-4}}{\tilde{\epsilon}^{d-4}}+
\begin{cases}
\mathcal{C}_{\mathcal{D}}^{(d)} + \ldots & \qquad\qquad  d \quad \mathrm{odd}\\
& \mathrm{for}\\
\mathcal{C}_{\mathcal{D}}^{(d)} \log\left(\frac{2R}{\tilde{\epsilon}} \right) + \ldots & \qquad\qquad d >2 \quad \mathrm{even}
\end{cases}
\end{equation}
where the constants $c_{d-4}$ are non-vanishing only for $d\ge5$. As advertised, this expression exhibits the independence of the $\mathcal{C}_{\mathcal{D}^{(d)}}$ under rescalings of $\tilde{\epsilon}$. 

The $\mathcal{D}$ subindex in expression \eqref{eq:Ceq} stands for the fact that it is still necessary  to perform a background subtraction in order to obtain a finite result. The background subtraction $\left[\ldots\right]_{\mathcal{D}}$ is defined for any quantity $\mathcal{F}$ depending on the defect data $\mathcal{F}[n,g \mu_{F_I}]$ as
\begin{equation}\label{eq:backsub}
\left[\mathcal{F}\right]_{\mathcal{D}} \equiv \mathcal{F}[n,g \mu_{F_I}]-n\,\mathcal{F}[1,0].
\end{equation}
Note in particular that $\mathcal{F}[1,0]$ is the vacuum result. As will be explained, $g \mu_{F_I}$ are the flavour monodromy sources and $n$ is the conical deficit/excess that characterises the defect.

In the prescription employed in \cite{Jensen:2013lxa,Conti:2025wwf} one first brings the metric \eqref{eq:gAnz},
\begin{equation}
d s^2_{d+1} = e^{2V} \left( \frac{ds^2(\text{Mink}_{1,d-2}) + dZ^2}{Z^2} \right) + h^2 d \varphi^2 + f^2 d y^2
\end{equation}
to the canonical Fefferman-Graham (FG) form\footnote{In \eqref{eq:asymptd} we have also allowed for the possibility that a conical defect, parametrised by the constant $n$, occurs at the boundary. We will see that this introduces a non-trivial monodromy for the gauge field associated to the $U(1)$ R-symmetry.},
\begin{equation}\label{eq:asymptd}
ds^2_{d+1} = L^2\frac{d\zeta^2}{\zeta^2}
+\frac{L^2}{\zeta^2}\Bigg[g_1\left(\frac{\zeta}{r_\perp}\right) ds^2(\text{Mink}_{1,d-2})\\
+ g_2\left(\frac{\zeta}{r_\perp}\right) dr_\perp^2+g_3\left(\frac{\zeta}{r_\perp}\right)r_\perp^2n^2 d\varphi^2\Bigg],
\end{equation}
where the functions $g_i$ behave to leading order as
\begin{equation}
g_i = 1+\mathcal{O}\left(\frac{\zeta}{r_\perp}\right).
\end{equation}
One then implements a holographic cut-off on the radial coordinate $\zeta$, $\Lambda_\zeta= \epsilon $. 
The coordinate transformation that brings the metric to this FG form was given in \cite{Jensen:2013lxa}, and reads
\begin{equation}\label{eq:FG1}
\zeta = Z\,G(y), \qquad r_\perp = Z\, F(y)
\end{equation}
where
\begin{equation}\label{eq:FG2}
G = e^{-\frac{1}{L}\int^y d y'f\sqrt{1-L^2e^{-2V}}}\qquad \mathrm{and} \qquad F = e^{-L\int^y dy'f\frac{e^{-V}}{\sqrt{e^{2V}-L^2}}}.
\end{equation}
Inverting this coordinate transformation one ultimately finds that the holographic cut-off in these solutions is realised as a particular constant $y$ surface.
Substituting in \eqref{eq:Ceq} and performing the background subtraction \eqref{eq:backsub} one then obtains the cut-off independent quantity that we associate with the defect contribution to the entanglement entropy.

\section{Co-dimension 2 monodromy defects in 6d (1,0) CFTs}\label{6dtheories}

In this section we study two different classes of solutions that we interpret as describing co-dimension 2 monodromy defects within 6d (1,0) CFTs living on NS5-D6-D8 brane intersections \cite{Brunner:1997gf,Hanany:1997gh}. The ambient CFTs are thus dual to the class of $\text{AdS}_7\times S^2\times I$ solutions to massive Type IIA supergravity constructed in \cite{Apruzzi:2013yva}, as shown in \cite{Gaiotto:2014lca,Cremonesi:2015bld}. We start by studying a solution to 7d U(1) gauged supergravity which we uplift to massive Type IIA supergravity, where it gives rise to a class of solutions that we interpret as describing 4d co-dimension 2 monodromy defects within the 6d (1,0) CFTs. We propose a brane set-up underlying the solutions and give the first steps towards identifying the dual 4d defect CFTs. Finally, we present a second class of solutions that we interpret as describing 2d co-dimension 2 monodromy defects within 6d (1,0) CFTs compactified on a hyperbolic plane.

\subsection{The $\text{AdS}_5\times S^1\times I$ monodromy defect}\label{7dsolution}

In this subsection we study a solution to 7d $U(1)$ gauged supergravity  that we interpret as describing 4d co-dimension 2 monodromy defects within 6d (1,0) CFTs. The solution is locally equivalent to the solution to 7d $U(1)^2$ gauged supergravity constructed in \cite{Ferrero:2021wvk}, describing M5-branes wrapped on a spindle, truncated such that it becomes a solution to 7d $U(1)$ gauged supergravity\footnote{Note that in this case a spindle interpretation is not possible.  The reason is well explained in \cite{Ferrero:2021wvk}: in order to have a spindle, we need the two integration constant $q_1$ and $q_2$ in the $7D$ $U(1)^2$ model to satisfy $q_1 q_2 \leq 0$. In our truncation we take $q_1 = q_2$, which clearly does not satisfy the above constraint.}. However, globally our solution contains a semi-compact $S^1\times I$ space instead of a spindle, which ultimately allows for the defect interpretation. In the following subsection we uplift this solution to massive Type IIA supergravity. 

The Lagrangian of 7d $U(1)$ gauged supergravity reads \cite{Cvetic:1999xp} 
\begin{equation}\label{eq:7DU1Lagrangian}
e^{-1} \mathcal{L} = R - \frac{1}{2} (\partial \eta)^2 - \frac{g^2}{2} V -\frac{1}{4} X^{-2} F^2,
\end{equation}
where
\begin{equation}
X = e^{-\frac{1}{\sqrt{10}} \eta}, \qquad F=dA,
\end{equation}
and the scalar potential is given by
\begin{equation}
V = - 4 X^2 -4 X^{-3} + \frac{1}{2} X^{-8}.
\end{equation}

The solution we are interested in here can be written 
\begin{equation} \label{eq:D7mingaugedsugrasolution}
\begin{split}
ds^2_7 & = \frac{2 y^{1/5} h^{2/5}}{g^2} \left[d s^2(\text{AdS}_5) +\frac{y}{P}d y^2+\frac{P}{h^2} n^2 d \varphi^2 \right], \\[2mm]
A & = \left(\alpha_0 - \frac{2^{2} n}{g}  \frac{y^2}{h} \right) d\varphi , \qquad \varphi\in [0,2\pi], \qquad X = \frac{y^{2/5}}{h^{1/5}}.
\end{split}
\end{equation}
Here $d s^2(\text{AdS}_5)$ is a unit radius metric, and the functions $h$ and $P$ depend only on $y$, 
\begin{equation}
h = y^2+q, \qquad P = h^2-4 y^3 ,
\end{equation}
with $q$ a constant. This solution is obtained setting $q_1=q_2=q$ in the solution to 7d $U(1)^2$ gauged supergravity constructed in \cite{Ferrero:2021wvk},  to which it is locally equivalent, and preserves the same number of supersymmetries. However, we take the $y$ coordinate to live in the $[y_{core},\infty)$ interval, which makes possible the defect interpretation \cite{Gutperle:2022pgw}\footnote{As a clarification, note that the solution constructed in \cite{Gutperle:2023yrd}, which was obtained setting  $q_2 = 0$ in the solution in \cite{Gutperle:2022pgw}, exhibits an enhancement of the supersymmetry to $\mathcal{N}=2$, which allowed the use of the electrostatic formulation of the LLM class of solutions \cite{Lin:2004nb} to explicitly describe the 4d defect CFT dual to the solution. One can further check that the solution in \cite{Gutperle:2023yrd}  is not a solution to 7d $U(1)$ gauged supergravity.}. 
As in \cite{Ferrero:2021wvk} we have set to zero the non-extremality parameter present in the solutions \cite{Lu:2003iv,Liu:1999ai}, from where the spindle constructions in  \cite{Ferrero:2021wvk} were obtained upon double analytical continuation. In this way we deal with a supersymmetric solution, preserving 4 real supercharges. The defect interpretation that follows holds straightforwardly to the case of non-vanishing non-extremality parameter.

The defect interpretation of the solution \eqref{eq:D7mingaugedsugrasolution} relies in part on the fact that in the limit $y\rightarrow\infty$ the $\text{AdS}_7$ vacuum, deformed by a non-vanishing monodromy for the gauge field $A$, emerges asymptotically. The AdS$_7$ vacuum with radius $L_{\text{AdS}_7} = \frac{2 \sqrt{2}}{g}$, in its parametrisation as a foliation by AdS$_5$ and S$^1$ times an interval, 
is obtained from the general solution \eqref{eq:D7mingaugedsugrasolution} when $q=0$ and $n=1$, 
\begin{equation}\label{AdS4}
ds^2(\text{AdS}_7)  = \frac{2}{g^2} \left( y\, ds^2(\text{AdS}_5) + \frac{1}{(y-4) y} dy^2 + (y-4) d \varphi^2  \right), \qquad A=0, \qquad X=1.
\end{equation}

In turn, $y_{core}$, the location of the ``core'' of the solution, is taken as the outermost root of the quartic polynomial $P$. Expanding near the core and substituting $\rho = 2 \sqrt{y-y_{core}}$, the metric of the $S^1\times I$ space spanned by $\varphi$ and $y$ takes the form
\begin{equation}\label{eq:7dmingaugedexpansion}
ds^2_{S^1\times I} = \frac{4 y_{core}}{P'}\left( d\rho^2 + \frac{(P')^2 n^2}{2^4 y_{core}^4 }\rho^2 d\varphi^2 \right),\\[2mm]
\end{equation}
that implies one must take
\begin{equation}\label{eq:7dU1regularitycondition}
n = \frac{4 y_{core}^2}{P'}\\[2mm]
\end{equation}
for the metric to be regular at the core. Substituting now the highest root of $P$ in \eqref{eq:7dU1regularitycondition}, solving for $q$ and plugging the result back in $y_{core}$ one gets
\begin{equation}
\label{eq:ycore7Dmingaugedsugra}
y_{core} = \frac{(1+3n)^2}{4 n^2}, \qquad q = \frac{(n-1) (1+3 n)^3}{16 n^4}.
\end{equation}

In \eqref{eq:D7mingaugedsugrasolution} the gauge parameter $\alpha_0$ is fixed by the condition that the gauge field is non-singular at $y=y_{core}$,
\begin{equation}\label{Acore}
A(y_{core}) = 0.
\end{equation}
This gives
\begin{equation}\label{alpha0}
\alpha_0 = \frac{1+ 3n}{g}.
\end{equation}
Thus, the gauge field at infinity reads to leading order
\begin{equation}
A(y \to \infty) = \frac{1-n}{g} d\varphi.
\end{equation}
This gives rise holographically to a non-trivial monodromy for the $U(1)$ R-symmetry gauge field, $g \mu_R = 1 - n$. Therefore the way to implement a non-trivial monodromy for this solution is to allow for the presence of a conical singularity.  In turn, the metric becomes at infinity 
\begin{equation} \label{eq:D7mingaugedsugrasolutionasymptotic}
ds^2_7(y \to \infty)= \frac{2}{g^2} \left[y\,d s^2(\text{AdS}_5) +y n^2 d \varphi^2+ \frac{1}{y^2}dy^2 \right], \qquad X = 1,
\end{equation}
thus exhibiting a deformation of the AdS$_7$ vacuum \eqref{AdS4} induced by the monodromy. Going to Fefferman-Graham (FG) coordinates through the mapping 
\begin{equation}\label{eq:FG7d}
y(\rho) = \rho^2 + 2 + \frac{5-2q}{5}\frac{1}{\rho^2}-\frac{4q}{15}\frac{1}{\rho^4}+\ldots,
\end{equation}
the metric near the AdS$_7$ boundary becomes
\begin{equation}\label{FGmetric}
ds^2_7(\rho\rightarrow\infty) = \frac{L^2}{\rho^2}d\rho^2 + \frac{L^2}{4}\rho^2\left[ds^2(AdS_5)+n^2d\varphi^2 + \ldots \right],
\end{equation}
as the FG radius $\rho$ approaches infinity.

Via the standard holographic dictionary, the term in square brackets in \eqref{FGmetric} is the metric of the boundary theory, that we can write
\begin{equation}\label{deformAdS7}
ds^2(AdS_5) + n^2 d\varphi^2 = \frac{1}{\zeta^2}\left(-dt^2 + d\vec{x}^2 +d\zeta^2 + n^2 \zeta^2 d\varphi^2 \right)
\end{equation}
where $\zeta$ is the AdS$_5$ radial coordinate. Thus, if we Weyl transform to $\mathbb{R}^{1,5}$, we see that the boundary exhibits a conical defect with deficit (or excess) angle $2\pi(1-n)$.

\subsection{Uplift to massive Type IIA}\label{upliftmassiveIIA}

Since \eqref{eq:D7mingaugedsugrasolution} is a solution to 7d $SU(2)$ minimal gauged supergravity we can use the consistent truncation built in \cite{Passias:2015gya} to uplift it to massive Type IIA supergravity. In our case we just have one gauge field $A$, so in the conventions of  \cite{Passias:2015gya} $A^i = (0,0,A)$. Performing this uplift a new class of solutions to massive Type IIA supergravity is obtained which are foliations of the 7d space and a deformed $S^2$ over an interval, parametrised by a new coordinate $z$. The solutions in this class are specified by a function $\alpha(z)$ that satisfies a Bianchi identity dependent on the Romans mass
\begin{equation}
\alpha'''=-2\pi^3 3^4 F_0.
\end{equation}
They read
\begin{align}\label{7duplift}
\frac{ds^2_{10}}{\sqrt{2} \pi} & = g^2  \sqrt{-\frac{\alpha}{\alpha''}} X^{-1/2} ds^2_7 + X^{5/2} \sqrt{-\frac{\alpha''}{\alpha}} \left( dz^2 + \frac{\alpha^2}{{\alpha'}^2-2\alpha \alpha'' X^5 } Ds^2(\text{S}^2) \right),
\end{align}
where $ds^2_7$ and $X$ are the metric and scalar in \eqref{eq:D7mingaugedsugrasolution}, $Ds^2(\text{S}^2)$ is given by
\begin{equation}\label{deformS2}
Ds^2(\text{S}^2) = d\theta^2 + \sin^2\theta (d\phi - g A)^2 \ ,
\end{equation}
$(\theta,\phi)$ parametrise the $S^2$ and
\begin{equation}\label{deformS2bis}
\text{vol}_2=\sin{\theta}d\theta\wedge (d\phi-gA),
\end{equation}
where $A$ is the gauge field in \eqref{eq:D7mingaugedsugrasolution}. 
The dilaton and fluxes are given by 
\begin{equation}\label{7dupliftfluxes}
\begin{split}
e^{\Phi} & = 2^{\frac{5}{4}} 3^4 \pi^{5/2}   \frac{X^{5/4}}{({\alpha'}^2-2\alpha\alpha'' X^5)^{1/2}} \Bigl(-\frac{\alpha}{\alpha''}\Bigr)^{3/4}, \qquad F_0 =-\frac{\alpha'''}{2 \pi^3 3^4}  , \\[2mm]
B_2 & = \pi \left(\frac{\alpha \alpha'}{{\alpha'}^2-2\alpha \alpha'' X^5} \, \text{vol}_2 - z \left(\text{vol}_2  + g \cos \theta F \right) \right) , \\[2mm]
F_{2} & = \frac{1}{2 \pi^2 3^4} \left(\alpha'' \left(\text{vol}_2 + g \cos \theta F \right) - \frac{\alpha \alpha' \alpha'''}{{\alpha'}^2-2\alpha\alpha'' X^5} \text{vol}_2  \right), \\[2mm]
F_4 & = \frac{ g }{2 \pi 3^4}  \left( \frac{\alpha\alpha' \alpha''}{{\alpha'}^2-2\alpha\alpha'' X^5} \cos \theta F \wedge \text{vol}_2 + \alpha'' \, \sin^2 \theta dz \wedge F \wedge d \phi \right).
\end{split}
\end{equation}
Finally, the Page fluxes read
\begin{eqnarray}\label{Pagefluxes}
&&\hat{F}_2=\frac{1}{2\pi^2 3^4}(\alpha''-z\alpha''')(\text{vol}_2+g\cos{\theta}F), \\
&&\hat{F}_4=\frac{1}{3^4\pi}g\cos{\theta}(-\alpha'+\alpha'' z-\frac12 \alpha'''z^2)\text{vol}_2\wedge F+\frac{1}{2\pi 3^4}d\Bigl(g\alpha'\sin^2{\theta}F\wedge d\phi\Bigr).\nonumber
\end{eqnarray}

The uplift to massive Type IIA guarantees that these solutions asymptote locally to the class of $\text{AdS}_7\times S^2\times I$ solutions constructed in \cite{Apruzzi:2013yva}. As in 7d, the $\text{AdS}_7$ subspace is deformed by the non-trivial monodromy of the gauge field. In turn, the volume of the squashed 2-sphere becomes at $y\rightarrow\infty$
\begin{equation}
\text{vol}_2=\sin{\theta}d\theta\wedge (d\phi + (n-1) d\varphi).
\end{equation}
Defining 
\begin{equation}
{\tilde \phi}=\phi + (n-1) \varphi,
\end{equation} 
${\tilde \phi}$ has standard period $2\pi$ if 
\begin{equation}\label{quanti}
n-1 \in \mathbb{Z},
\end{equation}
giving rise to well-defined NS5 and D6 quantised charges upon integration of $B_2$ and $\hat{F}_2$ on the 2-sphere spanned by $(\theta,\tilde{\phi})$ located at $y=\infty$. For finite $y$ we can define the connection
\begin{equation}
{\tilde A}= -\cos{\theta}(d\phi-gA),
\end{equation}
such that 
\begin{equation}\label{tildeF}
 {\tilde F}=d{\tilde A}=\text{vol}_2+g\cos{\theta}F,
\end{equation}
from which the Page fluxes can be written, 
\begin{eqnarray}
&&\hat{F}_2=\frac{1}{2\pi^2 3^4}(\alpha''-z\alpha'''){\tilde F}, \\
&&\hat{F}_4=\frac{1}{2\pi 3^4}(-\alpha'+\alpha'' z-\frac12 \alpha'''z^2){\tilde F}\wedge {\tilde F} +\frac{1}{2\pi 3^4}d\Bigl(g\alpha'\sin^2{\theta}F\wedge d\phi\Bigr). \label{F4flux}
\end{eqnarray}
The associated quantised charges will be computed in subsection \ref{field-theory}. Before that we turn to the computation of the defect entanglement entropy.

\subsubsection{Defect entanglement entropy}\label{holocc1}

In this subsection we compute the defect entanglement entropy associated to the class of solutions \eqref{7duplift}-\eqref{7dupliftfluxes}. As recalled in section \ref{subsec:dEE}, the first step is to compute the quantities $\mathcal{C}^{(6)}[n,\mu_{F_I}]$ and $\mathcal{C}^{(6)}[1,0]$, substituting in equation \eqref{eq:backsub}. For the class of solutions \eqref{7duplift}-\eqref{7dupliftfluxes} there is just the monodromy associated to the R-symmetry, so we can simply write $\mathcal{C}^{(6)}[n,g \mu_{F_I}]= \mathcal{C}^{(6)}[n]$. Moreover, the integral in \eqref{eq:Ceq} simplifies considerably, giving 
\begin{equation}\label{calC1}
\mathcal{C}^{(6)}[n] = - \frac{n \pi^2}{32 G_N^{(7)}} \int_{y_{core}}^{\infty} d y\, y \,.
\end{equation}
As already pointed out this integral diverges. In order to cure the divergence we implement the cut-off in FG coordinates, $\zeta=\epsilon$, and use the change of coordinates \eqref{eq:FG1}-\eqref{eq:FG2} to obtain the cut-off for $y$\cite{Jensen:2013lxa,Conti:2025wwf}. We find
\begin{equation}\label{eq:Lambda7d}
\Lambda = \frac{Z^2}{\epsilon^2} - \frac{2}{5} q \frac{\epsilon^2}{Z^2} - \frac{3}{50}q^2 \frac{\epsilon^6}{Z^6} + ...
\end{equation}
where the ellipsis are subleading contributions that will not enter the computations. 
The regularised expression for \eqref{calC1} then reads
\begin{equation}
\begin{split}\label{eq:intdiverge}
{\cal{C}}^{(6)}[n] & = - \frac{n \pi^2}{32 G_N^{(7)}}  \int_{y_{core}}^{\Lambda} d y\, y \\[2mm]
& = - \frac{n \pi^2}{64 G_N^{(7)}} \left[ \left( \frac{Z^2}{\epsilon^2} - \frac{2}{5} q \frac{\epsilon^2}{Z^2}\right)^2 - y_{core}^2 \right].
\end{split}
\end{equation}
At this point we implement the background subtraction introduced in \eqref{eq:backsub}, where 
\begin{equation}
{\cal{C}}^{(6)}[1] ={\cal{C}}^{(6)}[n=1] = - \frac{\pi^2}{64 G_N^{(7)}} \left[\frac{Z^4}{\epsilon^4} - 16 \right].
\end{equation}
Substituting in \eqref{eq:backsub} we finally find
\begin{equation}\label{eq:defEEAdS5IIA}
\begin{split}
{\cal{C}}_D^{(6)} & = \frac{\pi^2}{5 \, 2^{10}} \, \frac{(1-n) \left(1 + 29 n + 227 n^2 +767 n^3 \right)}{n^3} \, a^{(1,0)},
\end{split}
\end{equation}
where $a^{(1,0)}$ is the Weyl anomaly of the 6d (1,0) CFT, in the conventions of \cite{Cremonesi:2015bld}\footnote{In our conventions, the ten dimensional Newton constant is $G_N = 8 \pi^6$.}
\begin{equation}\label{eq:ccAdS7vacuum}
a^{(1,0)}= \frac{2^9 \pi^2}{3^5 7 G_N} \int dz (-\alpha \alpha'').
\end{equation}

\subsection{Towards a field theory interpretation}\label{field-theory}

In this subsection we give the first steps towards identifying the 4d CFTs living on the worldvolume of the defect described holographically by the solutions \eqref{7duplift}-\eqref{7dupliftfluxes}. We propose a 4d Hanany-Witten brane set-up that preserves the same number of supersymmetries as the solutions and consists of D4 defect branes embedded within the NS5-D6-D8 brane intersection where the ambient theory lives. 
As a probe of this set-up, we also construct the baryon vertex of the defect theory, which we compare to that of the ambient CFT.

\subsubsection{Brane set-up}

In this subsection we propose a brane set-up underlying the $\text{AdS}_5$ foliated class of solutions \eqref{7duplift}-\eqref{7dupliftfluxes}. 

We start  by looking at the branes of the ambient theory, starting with the NS5-branes. From the $B_2$-field given by  \eqref{7dupliftfluxes} one sees that a NS5-brane is created at $z=k$ upon integration on the 2-sphere spanned by $(\theta,{\tilde \phi})$ located at $y=\infty$, since
\begin{equation}\label{Q5first}
Q_5^k=\frac{1}{4\pi^2}\int H_3=\frac{1}{4\pi^2}\oint_{S^2_\infty}\Bigl(B_2(z=k+1)-B_2(z=k)\Bigr)=1.
\end{equation}
This is equivalent to imposing that $B_2$ lies in the fundamental region, namely, 
\begin{equation}
\frac{1}{4\pi^2}\oint_{S^2_\infty}B_2\in [0,1],
\end{equation}
which in turn implies that a large gauge transformation with gauge parameter $k$ must be performed at each $z\in [k,k+1]$ interval, restricting $B_2$ to the fundamental region. Therefore, we take
\begin{equation}
B_2 =  \pi \left(\frac{\alpha \alpha'}{{\alpha'}^2-2\alpha \alpha'' X^5} \, \text{vol}_2 - (z-k) {\tilde F}  \right) \qquad \text{for} \qquad z\in [k,k+1],
\end{equation}
with ${\tilde F}$ as in \eqref{tildeF}.
Using this expression for $B_2$ one derives the 2-form Page flux
\begin{equation}
\hat{F}_2=\frac{1}{162\pi^2}(\alpha''-(z-k)\alpha'''){\tilde F} \qquad \text{for} \qquad z\in [k,k+1],
\end{equation}
from which the D6-brane charge in the $z\in [k,k+1]$ interval can be computed by integrating on $S^2_\infty$, 
\begin{equation}\label{Q6first}
Q_6^k=\frac{1}{2\pi}\oint_{S^2_\infty}\hat{F}_2=\frac{1}{81\pi^2}(\alpha''-(z-k)\alpha''').
\end{equation} 
In turn, the D8-brane charge is computed from the Romans mass, as
\begin{equation}
Q_8=2\pi F_0.
\end{equation}

Since both the defect and vacuum solutions are specified by the function $\alpha(z)$, that satisfies the Bianchi identity 
\begin{equation}\label{Bianchi}
\alpha'''=-162\pi^3 F_0,
\end{equation}
the most general solution can be constructed glueing local solutions with D8-branes, allowing for the Romans mass to jump along $z$. $\alpha(z)$ can then be taken as  a piecewise cubic function \cite{Cremonesi:2015bld,Nunez:2018ags}
\begin{equation}
\alpha_k(z)=-\frac{27}{2}\pi^2\beta_k(z-k)^3+\frac12\gamma_k (z-k)^2+\delta_k (z-k)+\mu_k \qquad \text{for} \qquad z\in [k,k+1],
\end{equation}
with $\beta_k$ integer numbers defined in each interval from the Romans mass at that interval,
\begin{equation}\label{Q8}
\beta_k=2\pi F_0^k=Q_8^k \qquad \text{for} \qquad z\in [k, k+1].
\end{equation}
In turn $(\gamma_k,\delta_k,\mu_k)$ are determined imposing continuity in $\alpha$, $\alpha'$ and $\alpha''$ at $z=k$. This is analogous to the way the most general $\text{AdS}_7\times S^2\times I$ solutions in \cite{Apruzzi:2013yva} were constructed.
Substituting  $\alpha_k(z)$ in \eqref{Q6first} one finds for the D6-brane charge at the $z\in [k,k+1]$ interval,
\begin{equation}\label{Q6second}
Q_6^k=\frac{\gamma_k}{81\pi^2}.
\end{equation}
The continuity condition of $\alpha_k''$ at $z=k$ implies that
\begin{equation}\label{gamma}
\gamma_k=\gamma_{k-1}-81\pi^2\beta_k,
\end{equation}
which guarantees that $Q_6^k$ is an integer number.
Together with the D8-brane charge given by \eqref{Q8} and the NS5-brane charge given by \eqref{Q5first} we thus find the quantised charges of the branes that underly the $\text{AdS}_7\times S^2\times I$ asymptotic geometry. The associated brane set-up, depicted in Table \ref{table1}, consists of NS5-branes located at $z=k$, D6 colour branes stretched between them, and orthogonal flavour D8-branes. The resulting configuration is $\frac12$-BPS with $SU(2)$ R-symmetry realised geometrically on the $S^2$.

\begin{table}[ht]
	\begin{center}
		\begin{tabular}{| l | c | c | c | c| c | c| c | c| c | c |}
			\hline		    
			& $x^0$ & $x^1$  & $x^2$ & $x^3$ & $x^4$ & $x^5$ & $x^6$ & $x^7$ & $x^8$ & $x^9$\\ \hline
			D6 & x & x & x & x & x &x  &x  &   &   &   \\ \hline
			NS5 & x & x &x  & x & x & x  &   &   &  &   \\ \hline
			D8 & x & x & x &x  & x &  x &  &x  & x & x  \\ \hline
		\end{tabular} 
	\end{center}
	\caption{$\frac12$-BPS brane intersection associated to the $\text{AdS}_7\times S^2\times I$ asymptotic geometry. $(x^0,x^1,x^2,x^3,x^4,x^5)$ parametrise the common 6d intersection. $x^6$ is the field theory direction, along which the D6-branes stretch.  $(x^7, x^8, x^9)$ parametrise the $\mathbb{R}^3$ where the $S^2$ lives. Note that according to the analysis in  \cite{Bobev:2016phc} $x^6$ and the radius of the $S^2$ mix in the near horizon limit to produce the radial direction of $\text{AdS}_7$ and the coordinate $z$.}
		\label{table1}	
\end{table} 

The same brane configuration depicted in Table \ref{table1} arises also in the defect solutions, with the branes now wrapped on the $\text{AdS}_5\times S^1$ submanifold of the asymptotically $\text{AdS}_7$ factor, as inferred from equation \eqref{deformAdS7}. Moreover, in this case there are also D4-branes associated to the RR 4-form Page flux given by \eqref{F4flux}. In particular integrating over the $S^1_{\tilde \phi}$ cycle located at $\theta=\pi/2$, the $S^1_{\varphi}$ cycle and the intervals $I_y=[y_{core},\infty)$, $I_z=[k,k+1]$ one gets 
\begin{equation}\label{Q4}
Q_4^k=\frac{1}{(2\pi)^3}\int \hat{F}_4=\frac{1}{4\pi^2 3^4}(\gamma_k-\frac{81}{2}\pi^2\beta_k)p,
\end{equation}
where $p\equiv n-1 \in \mathbb{Z}$, as implied by the condition \eqref{quanti}. Given \eqref{gamma} this can give an integer number for properly chosen $\{\beta_0,\dots,\beta_k\}$.

The picture that arises is that the holographic defect solutions are associated to the brane intersection depicted in Table \ref{table2}.
 \begin{table}[ht]
	\begin{center}
		\begin{tabular}{| l | c | c | c | c| c | c| c | c| c | c |}
			\hline		    
			& $x^0$ & $x^1$  & $x^2$ & $x^3$ & $x^4$ & $x^5$ & $x^6$ & $x^7$ & $x^8$ & $x^9$\\ \hline
			D6 & x & x & x & x & x &x  &x  &   &   &   \\ \hline
			NS5 & x & x &x  & x & x & x  &   &   &  &   \\ \hline
			D8 & x & x & x &x  & x &  x &  &x  & x & x  \\ \hline
			D4 & x & x & x & x &  &  &  & x  & & \\ \hline 
		\end{tabular} 
	\end{center}
	\caption{$\frac14$-BPS brane intersection associated to the $\text{AdS}_5$ solutions. $(x^0,x^1,x^2,x^3)$ are the directions where the 4d dual CFT lives. $(x^4,x^5)$ are the directions where the $SO(2)$ global symmetry is realised. $x^6$ is the field theory direction, along which the D6  stretch. $(x^8, x^9)$ are the directions realising the $SO(2)$ R-symmetry. As in the ambient theory, it is plausible that the direction 6 and the radial directions of the 45 and 89 directions mix in the near horizon limit to produce the radius of $\text{AdS}_5$, the radius of the asymptotic $\text{AdS}_7$  and the direction $z$.}	
	\label{table2}	
\end{table} 
This consists of a Hanany-Witten brane set-up containing the {\it ambient} branes associated to the $\text{AdS}_7\times S^2\times I$ asymptotic solutions plus additional D4 {\it defect} branes, giving rise to a common 4d intersection. The mixing between the two $S^1$'s spanned by $\phi$ and $\varphi$ in the internal geometry (see \eqref{deformS2},\eqref{deformS2bis}) reduces the R-symmetry group from $SU(2)$ to $U(1)$, rendering the 4d theory $\mathcal{N}=1$ supersymmetric. Note that as in the ambient theory it is expected that there is a mixing between the two radial directions in the $(x^4,x^5)$ and $(x^8,x^9)$ planes and the field theory direction $x^6$, giving rise to the $\text{AdS}_5$ and $\text{AdS}_7$ radii and the $z$-direction in the geometry. 

An interesting avenue to pursue with this information at hand is the construction of the quiver field theory that lives in the brane set-up. From the quiver the $a$ and $c$ central charges could be computed from the R-symmetry t'Hooft anomalies of the fermionic degrees of freedom, using the expressions
\begin{equation}
a(\epsilon)=\frac{3}{32}(3\text{Tr}R^3_\epsilon-\text{Tr}R_\epsilon), \qquad c(\epsilon)=\frac{1}{32}(9\text{Tr}R^3_\epsilon-5\text{Tr}R_\epsilon),
\end{equation}
where the R-symmetry is given by $R_\epsilon=R_0+\frac12 \epsilon {\cal F}$ and $R_0$ and ${\cal F}$ are the R-symmetry and flavour currents of the quiver field theory. Using $a$-maximisation one could then determine $\epsilon$ and the $a$ and $c$ central charges of the $\mathcal{N}=1$ fixed point, and check whether they agree with the Weyl anomaly of the 4d defect CFT that can be extracted from \eqref{eq:defEEAdS5IIA}. Indeed, it is known \cite{Conti:2025wwf,Chalabi:2021jud} that the defect entanglement entropy is a linear combination of the Weyl anomaly of the defect CFT and the conformal weight of the defect. These results could also be compared to the value obtained from the anomaly polynomial of the ambient theory. We hope to address these interesting issues in a future publication.  

In the next subsection we turn to the analysis of the baryon vertex of the $\text{AdS}_5$ solutions.
 
\subsubsection{The D2 baryon vertex}

An interesting configuration that we can study in the defect theory is the baryon vertex. Baryon vertices were used in  \cite{Conti:2024qgx}  to distinguish between AdS spaces interpreted as describing defects and those interpreted as dual to deconstructed higher dimensional CFTs. In the latter case the baryon vertex would exhibit the same size and energy of the baryon vertex of the ambient theory, while in the first case it would exhibit a genuine lower dimensional behaviour. 
In this subsection we pursue this study for the defect solutions \eqref{7duplift}-\eqref{7dupliftfluxes}. 

\begin{figure}[t!]
 	\centering
 	{{\includegraphics[width=12cm]{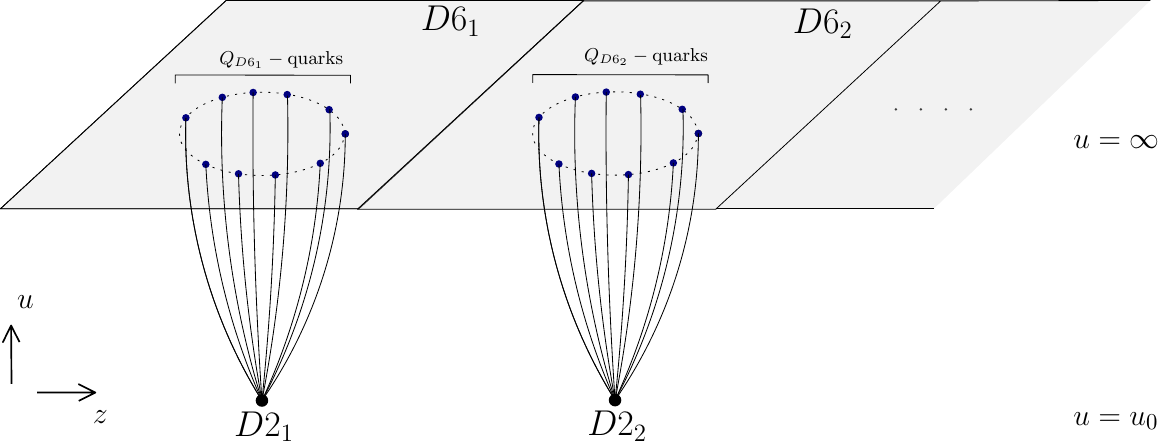} }}%
 	\caption{Baryon vertex configuration for the 6d ambient theory.}
 	\label{baryonvertex1}
 \end{figure}

First we recall that the baryon vertex associated to the $\text{AdS}_7\times S^2\times I$ solutions consists of D2-branes wrapped on the $S^2$ and located along the $z$-direction, in which fundamental strings stretched along the $\text{AdS}_7$ radial direction end \cite{Conti:2024qgx}. This is depicted in Figure \ref{baryonvertex1}. 

Similarly, the baryon vertex associated to the defect solutions contains D2-branes wrapped on the $S^2$ spanned by the $(\theta,{\tilde \phi})$ coordinates. This brane captures the $\hat{F}_2$ flux associated to the D6 colour branes, through the coupling
\begin{equation}
S^{WZ}_{D2}=T_2\int C_1\wedge F=T_2\oint_{S^2}\hat{F}_2\int dt A_t=Q_6 T_{F1}\int dt A_t.
\end{equation}
This coupling creates a tadpole that is cancelled with $Q_6$ fundamental strings stretched along the $\text{AdS}_5$ radial direction and ending on the brane. As explained in \cite{Conti:2024qgx}, since the gauge group of the field theory is a product of SU(N)'s whose ranks are the numbers of D6-branes stretched between the NS5-branes along the $z$ direction, a D2-brane located in each $z\in [k,k+1]$ interval is needed where the open strings can end, as depicted in Figure \ref{baryonvertex1}. Next we compute the size and energy of this configuration following closely \cite{Brandhuber:1998xy,Maldacena:1998im} (see also \cite{Conti:2024qgx}).

The action of the system consisting of the D2-brane in the $z\in [k,k+1]$ interval and the $Q_6^k$ fundamental strings ending on it is given by
\begin{equation}
S=S_{D2}+S_{Q_6^k F1}.
\end{equation}
The DBI action of the D2-brane reads
\begin{equation}
S_{D2}^{DBI}=-T_2\int dt\, \frac{2^{3/2}}{3^4}\sqrt{\frac{h}{y}}\sqrt{-\alpha_k\alpha''_k}\,u,
\end{equation}
in the parametrisation of $\text{AdS}_5$
\begin{equation}
ds^2_{\text{AdS}_5}=u^2dx_{1,3}^2+\frac{du^2}{u^2}.
\end{equation}
Here we have taken $\alpha'=0$ as in \cite{Conti:2024qgx}, as this minimises the energy. The Nambu-Goto action for the F1-strings is given by
\begin{equation}
S_{Q_6^k F1}=-Q_6^k\, T_{F1}\int dt d\sigma \, 2^{3/2}\pi \sqrt{h}\sqrt{-\frac{\alpha_k}{\alpha''_k}}\sqrt{\dot{u}^2+u^4},
\end{equation}
where we have parametrised the worldvolume coordinates by $(t,\sigma)$, the position in AdS$_5$ by $u=u(\sigma)$, and where the dot denotes derivative with respect to $\sigma$ (while a prime denotes derivative with respect to $z$). Substituting $Q_6^k$ as given by 
\eqref{Q6second}, where, we recall, $\gamma_k=\alpha''_k$ at $z=k$, we get
\begin{equation}\label{Q6strings}
S_{Q_6^k F1}=- T_{F1}\int dtd\sigma\,\frac{2^{3/2}}{3^4 \pi}\sqrt{h}\sqrt{-\alpha_k\alpha''_k}\sqrt{\dot{u}^2+u^4}.
\end{equation}
The WZ actions of the D2 and the F1-strings cancel each other as required by the tadpole cancelation condition. Following \cite{Brandhuber:1998xy,Maldacena:1998im,Conti:2024qgx} the equations of motion come in two sets, the bulk equation of motion for the F1-strings and the boundary equation of motion (as we are dealing with open strings) which contains as well a term coming from the D2-branes. The bulk equation of motion
\begin{equation}
\frac{\partial L_{F1}}{\partial u}-\frac{d}{d\sigma}\Bigl(\frac{\partial L_{F1}}{\partial \dot{u}}\Bigr)=0
\end{equation}
gives
\begin{equation}\label{bulk}
\frac{d}{d\sigma}\Bigl(\frac{u^4}{\sqrt{\dot{u}^2+u^4}}\Bigr)=0 \Rightarrow \frac{u^4}{\sqrt{\dot{u}^2+u^4}}=c(z).
\end{equation}
In turn, the equation of motion from the boundary is given by
\begin{equation}
\frac{\partial L_{D2}}{\partial u}=\frac{\partial L_{F1}}{\partial\dot{u}},
\end{equation}
which implies
\begin{equation}
\frac{\dot{u}_0}{\sqrt{\dot{u}_0^2+u_0^4}}=\frac{1}{2\sqrt{y}},
\end{equation}
where $u_0$ is the position of the D2-brane in $u$ and $\dot{u}_0=\dot{u}(u_0)$.
Combining this with \eqref{bulk} we get
\begin{equation}
\frac{u^4}{\sqrt{\dot{u}^2+u^4}}=u_0^2\sqrt{\frac{4y-1}{4y}}\equiv u_0^2\,\beta,
\end{equation}
for all D2-branes, independently on their positions in $z$. From here the size of the baryon is given by the usual expression in terms of the ${}_2F_1(a,b,c;x)$ hypergeometric function \cite{Brandhuber:1998xy}
\begin{equation}\label{ell}
\ell = \int_0^{\ell}d\sigma=\frac{\beta}{u_0}\int_{1}^{\infty}\frac{d\hat{u}}{\hat{u}^2\sqrt{\hat{u}^4-\beta^2}}=\frac{\beta}{3u_0}\,{}_2F_1(\frac12,\frac34,\frac74;\beta^2),
\end{equation}
where $\hat{u}=u/u_0$. As in the case of the $\text{AdS}_7$ solutions (see \cite{Conti:2024qgx}), the size of the vertex does not depend on its position in $z$. In the present (defect) case it depends however on its position in $y$. Notably, one can check that if the vertex is located at the value of $y_{core}$ associated to the vacuum $\text{AdS}_7$ solution, namely,  $y_{core}=4$, one recovers the size of the vertex of the $\text{AdS}_7$ solution obtained in \cite{Conti:2024qgx}.

The total on-shell energy of the vertex is given by
\begin{equation}
E=E_{D2}+E_{Q_6^k F1}=\frac{1}{3^4\sqrt{2} \pi^2}\sqrt{\frac{h}{y}}\sqrt{-\alpha_k\alpha''_k}\, u_0\Bigl(1+2\sqrt{y}\int_1^{\infty}d\hat{u}\frac{\hat{u}^2}{\sqrt{\hat{u}^4-\beta^2}}\Bigr).
\end{equation}
Contrary to our findings regarding the size of the vertex, the on-shell energy depends also on its position in $z$. Subtracting the energy of the constituents (F1-strings stretched from $u_0=0$ to infinity) the binding energy is given by
\begin{eqnarray}
E_{bin}&=&\frac{1}{3^4\sqrt{2} \pi^2}\sqrt{\frac{h}{y}}\sqrt{-\alpha_k\alpha''_k}\,u_0\Bigl(1+2\sqrt{y}\int_1^{\infty}d\hat{u}\Bigl[\frac{\hat{u}^2}{\sqrt{\hat{u}^4-\beta^2}}-1\Bigr]-2\sqrt{y}\Bigr)\nonumber\\
&=&-\frac{\sqrt{2}}{3^4\pi^2}\sqrt{h}\sqrt{-\alpha_k\alpha''_k}\, u_0\Bigl( {}_2F_1(\frac12,-\frac14,\frac34;\beta^2)-\frac{1}{2\sqrt{y}}\Bigr).
\end{eqnarray}
Once again one can check that the binding energy of the baryon vertex corresponding to the $\text{AdS}_7$ vacuum solution is recovered when the vertex is located at $y_{core}=4$. 

Finally, the binding energy per string is obtained dividing by the number of F1-strings. This gives
\begin{equation}\label{binding-string}
E_{bin(string)}=-\sqrt{2}\sqrt{h}\sqrt{-\frac{\alpha_k}{\alpha''_k}} \, u_0\Bigl( {}_2F_1(\frac12,-\frac14,\frac34;\beta^2)-\frac{1}{2\sqrt{y}}\Bigr).
\end{equation}
Combining this with \eqref{ell} we obtain
\begin{equation}
E_{bin(string)}=-\frac{f_k}{\ell}
\end{equation}
with
\begin{equation}
f_k=\beta \frac{\sqrt{2}}{3}\sqrt{h}\sqrt{-\frac{\alpha_k}{\alpha_k''}}\, {}_2F_1(\frac12,\frac34,\frac74;\beta^2)\Bigl({}_2F_1(\frac12,-\frac14,\frac34;\beta^2)-\frac{1}{2\sqrt{y}}\Bigr).
\end{equation}
One can check that $f_k$ is always positive. Therefore $dE_{bin(string)}/d\ell > 0$ and the configuration is stable.  Moreover, when $y_{core}=4$ the result for the baryon vertex configuration of the 6d ambient theory, found in \cite{Conti:2024qgx}, is recovered. Note that the behaviour with $1/\ell$ is the one dictated by conformal invariance \cite{Maldacena:1998im}. 
Finally, we would like to emphasise that the size and energy of the vertex associated to the defect solutions differ from those of the vertex of the  AdS$_7$ asymptotic geometry, in line with our defect interpretation\footnote{And in contrast to the solutions studied in \cite{Conti:2024qgx}, interpreted as deconstructed 6d theories.}.

\subsection{The $\text{AdS}_3\times S^1\times I \times H^2$ monodromy defect}\label{AdS3H2section}

In this subsection we construct a new solution to 7d $U(1)$ gauged supergravity that can be interpreted as describing co-dimension 2 monodromy defects within the 4d $\mathcal{N}=1$ CFTs obtained by compactifying the 6d (1,0) CFTs discussed in the previous subsections on a 2d hyperbolic plane, $H^2$. These CFTs are dual to the AdS$_5$ solutions to massive Type IIA supergravity constructed in \cite{Apruzzi:2015zna}, which are connected by RG flows to the $\text{AdS}_7\times S^2\times I$ solutions dual to the 6d (1,0) CFTs \cite{Merrikin:2022yho}. Analogously, one  may speculate that the uplift of our new $\text{AdS}_3\times S^1\times I \times H^2$ solution to massive Type IIA supergravity could be connected by an RG flow to the uplift of the $\text{AdS}_5\times S^1\times I $ solution constructed in subsection \ref{upliftmassiveIIA}. The solution that we present in this subsection is closely related to the solutions constructed in \cite{Boido:2021szx,Cheung:2022ilc,Suh:2022olh}, involving a two dimensional Riemann surface with constant curvature and a spindle, where the spindle has been replaced by a semi-compact space $S^1\times I$. As we will discuss, within the framework of 7d $U(1)$ gauged supergravity we will only be able to consider hyperbolic planes.

The possible connections between the two different classes of defect solutions constructed in this section are summarised in the following diagram
\[
\begin{tikzcd}
\text{AdS}_7 \times \text{S}^2 \times I  \arrow[d, "Defect"']  \arrow[r, "RG flow"]  & \text{AdS}_5 \times H^2 \times \tilde{\text{S}}^2 \times I \arrow[d, "Defect"] \\
 \text{AdS}_5 \times \text{S}^1 \times I \times \tilde{\text{S}}^2 \times I  \arrow[r, "RG flow"'] & \text{AdS}_3 \times \text{S}^1 \times I \times H^2 \times \tilde{\text{S}}^2 \times I
\end{tikzcd}
\]

The 7d solution that we discuss  is obtained by uplifting to 7d $U(1)$ gauged supergravity the solution to 5d minimal supergravity studied in \cite{Arav:2024exg} (see also \cite{Ferrero:2020laf}), interpreted as describing surface monodromy defects in 4d $\mathcal{N}=4$ SYM. The latter solution is given by
\begin{equation} \label{eq:5Dmin}
ds^2_5  = h \, ds^2(\text{AdS}_3)+ \frac{h}{4 P} dy^2 + \frac{P}{h^2} n^2 d\varphi^2 , \qquad A = n \left( \alpha_0 - \frac{q}{h} \right) d \varphi, \\[2mm]
\end{equation}
where
\begin{equation}\label{handP}
h = y + q, \qquad P = h^3 - y^2 \\[2mm]
\end{equation}
and $q$ is a real parameter.
$y_{core}$, the core of the solution, is taken to be the largest root of the cubic polynomial $P$. A similar analysis to the one performed in subsection \ref{7dsolution}, namely demanding that the 2d surface $\Sigma_2(y,\varphi)$ closes smoothly at $y_{core} $ and that the gauge field vanishes there, yields
\begin{equation}\label{ycoreandq}
y_{core} = \frac{(1 + 2 n)^3}{27 n^3} , \qquad q = \frac{(n-1) (2 n+1)^2}{27 n^3}, \qquad \alpha_0 = \frac{n-1}{3 n}. \\[2mm]
\end{equation}
For more details on this derivation see \cite{Conti:2025wwf}. \\

This solution can now be uplifted to 7d using the rules found in \cite{Faedo:2019cvr}, with the uplift being given by 
\begin{equation}
\begin{split}\label{eq:5Din7DSU2}
ds^2_7 & = \frac{8 X^8}{g^2} ds^2_5 + \frac{3}{2 X^2 g^2} ds^2(H^2), \\[2mm]
A_1& =A_2= 0, \qquad A_3 = \frac{3}{g} A + \frac{1}{g} \omega, \qquad X = \frac{3^{1/5}}{2^{2/5}}, \\[2mm]
{\cal{B}}_3 & = \frac{9}{2^{1/2} g^3} \star_5 dA - \frac{3}{\sqrt{2} g^3} A \wedge \text{vol}(H^2) + \frac{3}{\sqrt{2} g^3} dA \wedge \omega,
\end{split}
\end{equation}
where $\omega$ is a one-form satisfying\footnote{Here we have fixed the gauge coupling of the 5d theory to $g_{5d}=1$ for simplicity.} 
\begin{equation}
d \omega = - \text{vol}(H^2).
\end{equation}
It is easy to see that this solution asymptotes locally to the reduction to 7d of the AdS$_5 \times H^2$ solutions to massive Type IIA supergravity constructed in \cite{Apruzzi:2015zna}, performed in \cite{Passias:2015gya},
\begin{equation}
\begin{split}\label{eq:AdS5H27d}
ds^2_7 & = \frac{8 X^8}{g^2} ds^2(\text{AdS}_5) + \frac{3}{2 X^2 g^2} ds^2(H^2), \qquad A_3 =\frac{1}{g} \omega, \qquad X = \frac{3^{1/5}}{2^{2/5}}, \qquad {\cal{B}}_3 = 0,
\end{split}
\end{equation}
where the gauge field $A$ becomes pure gauge.

We can now uplift the solution \eqref{eq:5Din7DSU2} to massive Type IIA supergravity, using the uplift rules found in  \cite{Passias:2015gya}. One arrives at a new class of solutions, given by the same expressions \eqref{7duplift}-\eqref{7dupliftfluxes} with $A$  replaced by $A_3$ and two extra terms in the 4-form flux, that now reads
\begin{eqnarray}
F_4 & =& \frac{g }{2 \pi 3^4}  \bigg( \frac{\alpha\alpha'\alpha''}{{\alpha'}^2-2\alpha\alpha'' X^5} \cos \theta d A_3 \wedge \text{vol}_2 + \alpha'' \, \sin^2 \theta dz \wedge d A_3 \wedge ( d \phi - g A_3) \nonumber\\[2mm]
&& - 2 g \alpha'' X^4 dz \wedge \star_7 d {\cal{B}}_3 - \sqrt{2} g^2 d{\cal{B}}_3 \bigg).
\end{eqnarray}
These solutions preserve $\mathcal{N}=(0,2)$ supersymmetry in 2d. As the ambient theory is in this case identified with the IR CFTs obtained from 6d (1,0) CFTs compactified on a hyperbolic plane $H^2$, the new solutions can be interpreted as describing monodromy defects within the IR CFTs obtained from 6d (1,0) CFTs compactified on an $H^2$. Therefore it is
natural to speculate that such defect solutions could arise as the IR defect CFT of the 6d (1,0) defect CFT dual to the AdS$_5$ solutions constructed in subsection \ref{upliftmassiveIIA}, compactified on a hyperbolic plane. It would be interesting to try to make this interpretation precise.

Finally, we would like to mention that a generalisation to arbitrary two dimensional Riemann surfaces with constant curvature requires that the gauge group of the supergravity contains a $U(1)^2$ coupled to a three-form potential\footnote{This is clear at the level of the EOM. Another way to see it is to look at the solution \eqref{AdS3U(1)2}. In order to have a truncation to the $U(1)$ minimal subsector, we would need to set $F_1 = F_2$, that implies $z=0$. This sets $k=-1$ as it is clear from the warp factors.}. We carry out this extension in Appendix \ref{AdS3Riemann} in the context of 7d $U(1)^2$ matter-coupled gauged supergravity. Note however that the uplift of this gauged supergravity to massive Type IIA is not known.

\subsubsection{Defect entanglement entropy}

The entanglement entropy associated to the monodromy defects described by \eqref{eq:5Din7DSU2} can be computed performing an analogous calculation to the one presented in subsection \ref{holocc1}. In this case the ambient CFTs are the 6d (1,0) CFTs wrapped on $H^2$ studied in \cite{Apruzzi:2015zna}.

We find for the quantity ${\cal{C}}^{(4)}[n]$ 
\begin{equation}
{\cal{C}}^{(4)}[n] = \frac{n \pi}{2 G_N^{(3)}} \int_{y_{core}}^{\Lambda} dy,
\end{equation}
where the cut-off is now given by
\begin{equation}\label{eq:cutoff5dminimal}
\Lambda = \frac{Z^2}{\epsilon^2} - q - \frac{q}{2} \frac{\epsilon^2}{Z^2} + ... 
\end{equation}
Implementing the cut-off, we have
\begin{equation}
{\cal{C}}^{(4)}[n] = \frac{n \pi}{2 G_N^{(3)}} \left( \frac{Z^2}{\epsilon^2} - q - y_{core} \right)
\end{equation}
and
\begin{equation}
{\cal{C}}^{(4)}[1] = \frac{\pi}{2 G_N^{(3)}} \left( \frac{Z^2}{\epsilon^2}  - 1 \right).
\end{equation}
Performing the background subtraction \eqref{eq:backsub} to obtain a finite expression for the defect entanglement entropy,
\begin{equation}
{\cal{C}}^{(4)}_D = \frac{\pi}{18 n} (n-1) (1 + 5 n) a^{4d \times H^2},
\end{equation}
where we have substituted $y_{core}$ and $q$ as given in \eqref{ycoreandq} and in our conventions
\begin{equation}
a^{4d \times H^2} = \frac{3^3}{2^8 \pi} \text{vol}(H^2) a^{(1,0)},
\end{equation}
as obtained in \cite{Merrikin:2022yho}.

\section{Co-dimension 2 monodromy defects in the 5d Sp(N) fixed point theory}\label{5dtheories}

In this section we present similar results to those in the previous section for the case in which the ambient theory is the 5d Sp(N) fixed point theory \cite{Seiberg:1996bd}, dual to the $\text{AdS}_6\times S^3\times I$ solution to massive Type IIA supergravity constructed in \cite{Brandhuber:1999np}. Compared to the previous section an important difference is that an uplift to massive Type IIA supergravity of the 6d gauged supergravity with a larger, $U(1)^2$, gauge group is known, which will in practice allow us to construct more general defect solutions. Moreover, we will find $\text{AdS}_2\times S^1\times I\times \Sigma$ solutions to 6d $U(1)^2$ matter-coupled gauged supergravity where $\Sigma$ is a two dimensional Riemann surface with constant curvature or a spindle, and we will uplift them to massive Type IIA supergravity. These solutions extend the constructions  in \cite{Faedo:2021nub,Giri:2021xta,Suh:2022olh,Hristov:2024qiy,Couzens:2022lvg,Faedo:2022rqx}, involving a constant curvature 2d Riemann surface and a spindle or two spindles, to the case in which one of the 2d spaces has been replaced by a semi-compact $S^1\times I$ space, making possible the  defect interpretation.
Once uplifted to massive Type IIA supergravity we will interpret the solutions as describing defects within the 5d Sp(N) fixed point theory compactified on a Riemann surface or a spindle, after flowing to the IR. We will compute the defect entanglement entropy for the different defect solutions and provide strong evidence that the relation that exists for 1d, 2d and 4d defects between the defect entanglement entropy and a combination of the defect free energy/Weyl anomaly  and the conformal weight  may also hold for 3d monodromy defects in the 5d Sp(N) fixed point theory.

\subsection{The $\text{AdS}_4\times S^1\times I$ monodromy defect}\label{AdS4S1I6DU12}

In this subsection we reinterpret the solution to 6d $U(1)^2$ gauged supergravity constructed in \cite{Faedo:2021nub}, which contains a spindle, in terms of monodromy defects, by substituting the compact interval considered therein by a semi-compact one. This solution can be uplifted to massive Type IIA supergravity. The analogue of the 7d solution constructed in subsection  \ref{7dsolution} is a solution to 6d minimal supergravity that arises as a particular case, that we present in subsection \ref{6dminimal}.

The Lagrangian of 6d $U(1)^2$ gauged supergravity reads \cite{DAuria:2000afl,Faedo:2021nub}
\begin{equation} \label{eq:6DU12Lagrangian} 
{\mathcal{L}} = \sqrt{-g} \left( R - \mathcal{V} - \frac{1}{2} \partial \vec{\varphi}^2 - \frac{1}{2} \sum_{I=1}^2 (X^I)^{-2} F_I^2 \right) ,
\end{equation}
where $F_I= d A_I$ and the scalar fields $X^I$ can be parametrised in term of two real scalar fields $\vec{\eta}=(\eta_1,\eta_2)$ as
\begin{equation}
X_1 = e^{ - \frac{1}{\sqrt{2}} \eta_1 - \frac{1}{2 \sqrt{2}} \eta_2}, \qquad X_2 = e^{\frac{1}{\sqrt{2}} \eta_1 - \frac{1}{2 \sqrt{2}} \eta_2}.
\end{equation}
Introducing $X_0=(X_1 X_2)^{-3/2}$ the scalar potential takes the form
\begin{equation}
{\mathcal{V}} = m^2 X_0^2 - 4g^2 X_1 X_2 - 4mg \, X_0 (X_1 + X_2) ,
\end{equation}
where $g$ is the gauge coupling and $m$ the mass parameter. Finally, as in \cite{Faedo:2021nub} the gauge fields associated to the R and flavour symmetries are given by: 
\begin{equation}
A_R = A_1 + A_2, \qquad A_F = A_2 - A_1.
\end{equation} 

The solution we are interested in reads 
\begin{align}\label{eq:6DsolutionU12}
ds^2_6 & = \frac{3^{3/2}}{2^{3/2} g^{3/2} m^{1/2}} \left(y^2 h_1 h_2 \right)^{1/4} \left(ds^2(\text{AdS}_4) + \frac{y^2}{P} dy^2 + \frac{P}{h_1 h_2} n^2 d\varphi^2 \right), \nn \\[2mm]
A_I & = \left( \alpha_I - \sqrt{1- \frac{\nu}{q_I}} \frac{3}{2 g}\frac{y^3}{h_I} n \right) d\varphi , \qquad  \varphi\in [0,2\pi], \qquad X_I = \frac{3^{1/4} m^{1/4}}{2^{1/4} g^{1/4}}\frac{\left(y^2 h_1 h_2\right)^{3/8}}{ h_I}, \\[2mm]
h_I & = y^3 + q_I, \qquad P  = h_1 h_2 - y^4 - \nu y, \nn
\end{align}
where $q_I$ are two real parameters. $\nu$ is a non-susy parameter, that we will set to zero in order to focus on solutions preserving  ${\cal{N}}=2$ superymmetry in 3d. This solution is, locally, the same solution presented in  \cite{Faedo:2021nub}, with the extra addition of the conical parameter $n$, relevant for the defect interpretation. Our solution will however differ globally from the solution in \cite{Faedo:2021nub}, as explained below.

First, note that setting $n=1$ and $q_I=0$ the maximally supersymmetric AdS$_6$ vacuum with radius $L= \frac{2 g}{3}$ is recovered, 
in its parametrisation as a foliation by $\text{AdS}_4\times S^1$ over an interval\footnote{For $m=\frac{2 g}{3}$.},
\begin{equation}\label{eq:AdS4asAdS4}
ds_6^2 = \frac{9}{4 g^2} y^2 \left(ds^2(\text{AdS}_4) +  \frac{1}{y^2(y^2 -1)} dy^2+ \frac{ y^2 -1}{y^2} d\varphi^2 \right) , \qquad A_I=0, \qquad X_I=1.
\end{equation}
Instead, setting $q_1 = q_2 = q$ we obtain a solution to 6d $U(1)$ minimal supergravity, that we will further discuss in subsection \ref{6dminimal}. 

Let us now turn to the global analysis which procedes exactly as in the previous section. Taking $y$ in a finite interval and closing off the $S^1$ at both ends the spindle solution constructed in \cite{Faedo:2021nub} can be recovered. Instead, as we will now show taking $y$ in a semi-infinite interval $y\in [y_{core},\infty)$ the $\text{AdS}_6$ vacuum \eqref{eq:AdS4asAdS4}, deformed by non-vanishing monodromies for $A_I$, arises asymptotically, encouraging a defect interpretation. 
The gauge parameters $\alpha_I$ are fixed by the condition 
\begin{equation}\label{AI}
A_I(y_{core})=0,
\end{equation}
with $y_{core}$  taken as the highest root of the quartic polynomial $P$.
Expanding the metric of the $(y,\varphi)$ submanifold around $y_{core}$ and making the coordinate transformation $\rho=\sqrt{y - y_{core}}$, we obtain
\begin{equation}\label{eq:6DSigmacore}
ds^2_{(y,\varphi)}= \frac{4 y_{core}^2}{P'} \left( d\rho^2 + \left( \frac{P' n}{2 y_{core}^3} \right)^2 \rho ^2 d\varphi^2 \right).
\end{equation}
Therefore the metric is regular at $y_{core}$ if
\begin{equation}\label{eq:6Dregularity}
n = \frac{2 y^3}{P'} \bigg|_{y=y_{core}}.
\end{equation}
This gives an implicit definition of $y_{core}$. Indeed, imposing the conditions \eqref{AI} we find
\begin{equation}
\alpha_I = \frac{3 n y_{core}^3 }{2g (q_I + y_{core}) }.
\end{equation}
Using these quantities we get
\begin{equation}\label{twomonodromies}
A_I(\infty) = -\frac{3 n q_I }{2 g (q_I + y_{core}^3) } d\varphi .
\end{equation}
Substituting now in $ P = 0$ we find that its highest root is given by 
\begin{equation}
y_{core} = \frac{\sqrt{9 n^2 + g^2 \mu_R^2 - g^2 \mu_F^2 + 6 g \mu_R n }}{3 n},
\end{equation}
where we have defined
\begin{equation}\label{muRmuF}
A_1(\infty)+A_2(\infty) \equiv \mu_R\, d\varphi , \qquad A_2(\infty)-A_1(\infty) \equiv \mu_F\, d\varphi ,
\end{equation}
using the notation in \cite{Arav:2024exg,Arav:2024wyg}. Using now the regularity condition \eqref{eq:6Dregularity} we find the constraint
\begin{equation}
g \mu_R = 1-n,
\end{equation}
and we can finally express $y_{core}$ as
\begin{equation}\label{eq:ycore6DU12}
y_{core} =  \frac{\sqrt{( 1 + 2 n)^2 - g^2 \mu_F^2 }}{3 n }
\end{equation}
and
\begin{equation}\label{eq:qItomuF}
q_1 = \frac{\left((1 + 2 n)^2- g^2 \mu_{F}^2 \right)^{3/2} ( n-1+ g \mu_{F})}{27 n^3 (1 +2 n -g \mu_F )}, \qquad q_2 = \frac{\left((1 + 2 n)^2 - g^2 \mu_F^2 \right)^{3/2} (n-1- g \mu_{F})}{27 n^3 ( 1 +2 n + g \mu_F)}.
\end{equation}

Asymptotically, as $y\to\infty$ a metric that exhibits a deformation of the AdS$_6$ vacuum \eqref{eq:AdS4asAdS4} induced by the monodromies in \eqref{muRmuF} is obtained.
This is easily seen by moving to FG coordinates through the asymptotic coordinate transformation 
\begin{equation}\label{eq:6DU12FG}
y = \rho + \frac{9}{16 g^2} \frac{1}{\rho} - \frac{1}{8} (q_1 + q_2) \frac{1}{\rho^2} - \frac{9 (q_1 + q_2) }{2^6 5 g^2} \frac{1}{\rho^4} + \frac{(q_1^2 - 14 q_1 q_2 +q_2^2)}{2^8}\frac{1}{\rho^5} + ...
\end{equation}
giving rise to
\begin{equation}\label{FG6d1}
ds^2_6(\rho\rightarrow\infty) = L^2 \frac{d \rho^2}{\rho^2} + L^2 \frac{\rho^2}{4} \left( ds^2(\text{AdS}_4) + n^2 d\varphi^2 \right)+ ...
\end{equation}
where $L = \frac{3}{2 g}$. As per the standard holographic dictionary, the quantity in parentheses is to be identified with the boundary theory metric, which we can write
\begin{equation}\label{FG6d2}
ds^2(AdS_4) + n^2 d\varphi^2 = \frac{1}{\zeta^2}\left(-dt^2 + d\vec{x}^2 +d\zeta^2 + n^2 \zeta^2 d\varphi^2 \right),
\end{equation}
where $\zeta$ is the AdS$_4$ radial coordinate. Thus, if we Weyl transform to $\mathbb{R}^{1,4}$ we explicitly see that there is a deficit angle $2\pi (1-n)$. This is analogous to our analysis in subsection \ref{7dsolution}.

As discussed in  \cite{Faedo:2021nub}, the solution \eqref{eq:6DsolutionU12} is $\mathcal{N}=2$ supersymmetric in 3d. Two interesting limiting cases where we can obtain $\mu_F$ explicitly is when $q_1=q_2$ and when one of the $q_I$'s is set to zero.  In the $q_2=0$ case one has
\begin{equation}\label{eq:ycoreq20}
y_{core} = \sqrt{\frac{2+n}{3n}}, \qquad q_1 = 2 \frac{(n-1) \sqrt{2+n}}{(3n)^{3/2}}, \\[2mm]
\end{equation}
and
\begin{equation}
\alpha_1 = \frac{2 + n}{ 2 g}, \qquad \alpha_2 = \frac{3 n}{2 g}, \\[2mm]
\end{equation}
which imply that $A_2=0$.
Therefore, at $\rho=\infty$
\begin{equation}
A_1(\infty)= \frac{1-n}{g}d\varphi, 
\end{equation}
and one has
\begin{equation}\label{eq:muI6Dq2}
g \mu_R = 1 - n, \qquad g \mu_F = n-1.\\
\end{equation}
\\
We now turn to a more detailed analysis of the $q_1 = q_2$ case.

\subsection{The minimal $\text{AdS}_4\times S^1\times I$ monodromy defect}\label{6dminimal}

Another interesting particular case is when $q_1 = q_2\equiv q$ in the solution \eqref{eq:6DsolutionU12}, in which case we have a truncation from 6d $U(1)^2$ to 6d minimal supergravity (but no enhancement of supersymmetry). Thus, this solution can be considered the analogue in 6d of the solution in 7d constructed in section \ref{7dsolution}. The solution reads
\begin{equation}\label{eq:6dmingauged}
\begin{split}
ds^2_6 & = \frac{3^{3/2}}{2^{3/2} g^{3/2} m^{1/2}} \sqrt{yh}  \left(ds^2(\text{AdS}_4) + \frac{y^2}{P} dy^2 + \frac{P}{h^2}n^2  d\varphi^2 \right), \\[2mm]
A & = \left( \alpha_0 - \frac{3}{2 g}  \frac{y^3}{h} n \right) d\varphi , \qquad X = \frac{3^{1/4} m^{1/4}}{2^{1/4} g^{1/4}}\frac{y^{3/4}}{h^{1/4}}, \\[2mm]
h & = y^3 + q, \qquad  P  = h^2 - y^4. \\[2mm]
\end{split}
\end{equation}
One can check that in this case
\begin{equation}
y_{core}= \frac{1 + 2n}{3 n}, \qquad q = \frac{(n-1)(1+ 2 n^2)}{3^3 n^3}, \qquad \alpha_0= \frac{1 + 2n}{2 g}. \\[2mm]
\end{equation}
In the FG coordinates \eqref{eq:6DU12FG} we find \eqref{FG6d1} and \eqref{FG6d2}  for $\rho \to \infty$.
Since $A = A_1 = A_2 $, the asymptotic value of the gauge field at $\rho=\infty$ is given by
\begin{equation}
A(\infty)= - \frac{n-1}{2 g}d\varphi, \\[2mm]
\end{equation}
such that, using the definitions \eqref{muRmuF} we find
\begin{equation}\label{eq:muI6Dminimal}
g \mu_R = 1 - n, \qquad \mu_F = 0. \\[2mm]
\end{equation}

In the next subsection we present the uplift of this solution to massive Type IIA supergravity.

\subsection{Uplift to massive Type IIA}\label{6d-uplift-to10d}

The solution \eqref{eq:6DsolutionU12} can be uplifted to massive Type IIA supergravity using the rules presented in \cite{Faedo:2021nub} (see also \cite{Cvetic:1999xx}), to obtain a solution that is locally equivalent to the uplifted solution found in \cite{Faedo:2021nub} but with the spindle therein replaced by the $S^1\times I$ semi-compact space spanned by $\varphi$ and $y$. Namely,
\begin{align}
\label{eq:6DU12IIA}
d s^2_{10} & = \frac{\lambda^2}{\mu_0^{1/3} g^2} \Bigl[ \frac{9}{4} \frac{ \Delta_h^{1/2}}{y} \left( ds^2(\text{AdS}_4) +  \frac{y^2}{P} dy^2 + \frac{P}{h_1 h_2} n^2 d\varphi^2 \right) \nn \\[2mm]
& + \Delta_h^{-1/2} \left( y^3 d\mu_0^2 + h_1 \bigl(d\mu_1^2 + \mu_1^2 (d \phi_1 - g A_1)^2 ) + h_2 \bigl(d\mu_2^2 + \mu_2^2 (d \phi_2 - g A_2)^2\bigr) \right) \Bigr], \, \nn \\[2mm]
e^{\Phi} &= \lambda^2 \mu_0^{-5/6} y^{-3/2} \Delta_h^{1/4}, \qquad F_{0} = \frac{2g}{3\lambda^3} \,, \\[2mm]
F_{4} & = \frac{\lambda \, \mu_0^{1/3} h_1 h_2}{g^3 \Delta_h} \biggl[ \frac{U_h}{\Delta_h} \frac{\mu_1 \mu_2}{\mu_0} \, d \mu_1 \wedge d \mu_2 \wedge (d \phi_1 - g A_1) \wedge (d \phi_2 - g A_2) \nn \\[2mm]
& - g \sum_{I\neq J} F_I \wedge d\phi_J \wedge \Bigl( \mu_0 \mu_J d\mu_J - \frac{y^3}{h_J} \mu_J^2 \, d\mu_0 \Bigr) \nn \\[2mm]
& + \frac{y^3}{\Delta_h} \sum_{I\neq J}  \frac{h_J (h_I' - 3y^{-1} h_I)}{h_I} \mu_I^2 \, d y \wedge (d \phi_I - g A_I) \wedge (d \phi_J - g A_J) \wedge \Bigl( \mu_0 \mu_J d\mu_J - \frac{y^3}{h_J} \mu_J^2 \, d\mu_0 \Bigr) \biggr] \,. \nn
\end{align}
Here $\{\mu_0,\mu_1,\mu_2\}$ satisfy the constraint $\sum_{a=0}^{2}\mu_a^2=1$ and $I,J=\{1,2\}$, with the space spanned by $\{\mu_a,\phi_I\}$ parametrising a four-dimensional hemisphere. $\Delta_h$ and $U_h$ are defined as
\begin{equation}
\begin{split}
\Delta_h &= h_1 h_2\,\mu_0^2 + y^3 h_2\,\mu_1^2 + y^3 h_1\,\mu_2^2 \,, \\[2mm]
U_h &= 2 \bigl[ (y^3 - h_1)(y^3 - h_2) \mu_0^2 - y^6 \bigr] - \frac43 \Delta_h \,.
\end{split}
\end{equation}

Notably, one can verify that \eqref{eq:6DU12IIA} asymptotes locally to the $\text{AdS}_6\times S^3\times I$ solution to massive Type IIA supergravity found in \cite{Brandhuber:1999np}\footnote{Including the extra parameter $\lambda$ introduced in  \cite{Faedo:2021nub} to make the uplifted solution globally consistent.}, 
\begin{equation}
\begin{split}
\label{eq:AdS6vacuum}
\frac{ds^2_{10}}{\lambda^2} & =  \frac{1}{({\sin{\xi}})^{1/3} g^2} \, \biggl[\frac{9}{4} ds^2(\text{AdS}_6) +  \left( d\xi^2 + \cos^2 \xi \, ds^2(\text{S}^3) \right) \biggr] \,, \\[2mm]
e^\Phi & = \lambda^2 ({\sin{\xi}})^{-5/6} \,, \\[2mm]
F_{0} & = \frac{2 g}{3\lambda^3} \,,  \qquad  F_{4} = -\lambda \frac{10 \cos^3 \xi ({\sin{\xi}})^{1/3}}{3g^3} \, d\xi \wedge \text{vol}(\text{S}^3) \, ,
\end{split}
\end{equation}
in the parametrisation
\begin{equation}
\mu_0=\sin{\xi}, \qquad \mu_1=\cos{\xi}\sin{\eta},\qquad \mu_2=\cos{\xi}\cos{\eta},
\end{equation}
with $\xi\in [0,\pi/2]$, $\eta\in [0,\pi/2]$,
\begin{equation}
ds^2(S^3)=d\eta^2+\sin^2{\eta}d{\tilde \phi}_1^2+\cos^2{\eta}d{\tilde \phi}_2^2
\end{equation}
and
\begin{equation}
{\tilde \phi}_i=\phi_i+\frac{3n q_I}{2(q_I+y_{core}^3)}\varphi.
\end{equation}
If we use the expressions in \eqref{eq:qItomuF} this can be further simplified to
\begin{equation}
\tilde{\phi}_1 = \phi_1 + \frac{1}{2} (n-1+ g \mu_F ) \varphi, \qquad \tilde{\phi}_2 = \phi_2 + \frac{1}{2} (n-1 - g \mu_F ) \varphi.
\end{equation}
Note that we should impose the conditions ${\tilde \phi}_i\in [0,2\pi]$ to have a regular $S^3$ and to guarantee that the D4-brane charge is properly quantised.
As in the solutions discussed in the previous section the asymptotically $\text{AdS}_6$ subspace is deformed by the monodromies according to \eqref{FG6d2},
and there are extra fluxes proportional to them. The solution \eqref{eq:6DU12IIA} can thus be interpreted as describing 1/4-BPS three dimensional defects within the 5d $\mathcal{N}=1$ fixed point theory with gauge group $Sp(N)$ dual to the $\text{AdS}_6\times S^3\times I$ background \cite{Seiberg:1996bd,Brandhuber:1999np}. \\[2mm]

The uplift of the solution \eqref{eq:6dmingauged} can be obtained using the rules presented in \cite{Cvetic:1999un}, or, alternatively, setting $q_1=q_2\equiv q$ in \eqref{eq:6DU12IIA}. This gives
\begin{equation}\label{eq:6dmingaugedIIA}
\begin{split}
\frac{ds^2_{10}}{\lambda^2} & = \frac{\Delta^{1/2}}{({\sin{\xi}})^{1/3} X^{\frac{1}{2}}} \left[ ds^2_6 + \frac{ X^2}{g^2}  \left( d\xi^2 + \frac{\cos^2 \xi}{\Delta X^3} \left( \frac{1}{4} ds^2(\text{S}^2) + \left(ds^2(\text{S}^1) + \cal{A}  \right)^2 \right) \right) \right], \\[2mm]
e^{\Phi} & = \lambda^2 ({\sin{\xi}})^{-5/6}  X^{-5/4} \Delta^{1/4}, \qquad F_0 =  \frac{2 g}{3\lambda^3}, \\[2mm]
F_4 & = -\frac{2 \lambda }{3 g^3 \Delta^2} U ({\sin{\xi}})^{1/3} \cos ^3 \xi d \xi \wedge \text{vol}(\text{S}^3) -\frac{4 \lambda  ({\sin{\xi}})^{4/3} \cos ^4 \xi }{g^3 X^3 \Delta^2} dX \wedge \text{vol}(\text{S}^3) \\[2mm]
& + \frac{\lambda  ({\sin{\xi}})^{1/3} \cos \xi }{\sqrt{2} g^2} (F \wedge {\cal{A}} + F \wedge \text{vol}(\text{S}^1 ))\wedge d \xi - \frac{\lambda ({\sin{\xi}})^{4/3} \cos ^2 \xi }{ 4 \sqrt{2}  g^2 X^3 \Delta } F \wedge \text{vol}(\text{S}^2),
\end{split}
\end{equation}
where we now write the $S^3$ as a Hopf fibration, with $(\theta,\phi)$ parametrising the $S^2$ and
\begin{equation}
{\cal{A}} = \frac{1}{2} \cos \theta d \phi - \frac{g}{\sqrt{2}} A.
\end{equation}
In these expressions  $\Delta$ and $U$ are defined as 
\begin{equation}
\begin{split}
\Delta & = X \cos^2 \xi + \frac{\sin^2 \xi}{X^3}, \qquad U = \frac{\sin ^2 \xi }{X^6}-3 \cos ^2 \xi X^2 + \frac{4 \cos ^2 \xi}{X^2}-\frac{6}{X^2}.
\end{split}
\end{equation}
As in the solution \eqref{eq:6DU12IIA}, this solution asymptotes locally to the Brandhuber-Oz solution when $y\rightarrow\infty$, with extra fluxes and a deformation of $\text{AdS}_6$ due to the monodromy. 

We compute the defect entanglement entropy of these solutions in the next subsection.

\subsubsection{Defect entanglement entropy}

In order to obtain the defect entanglement entropy we proceed as in the previous examples, we first compute the quantity ${\cal{C}}^{(5)}[n, g \mu_F]$, where now we have also a contribution from the flavour monodromies. We find
\begin{equation}\label{CnmuF}
{\cal{C}}^{(5)}[n, g \mu_F] = -\frac{81 \pi^2 n}{16 g^4 G_N^{(6)}} \int_{y_{core}}^{\Lambda} dy \, y^2.
\end{equation}
Next we obtain the cut-off in the canonical form, 
\begin{equation}\label{cut-off6dU(1)2}
\Lambda = \frac{Z}{\epsilon} - \frac{q_1 + q_2}{8} \frac{\epsilon^2}{Z^2} - \frac{5}{128} (q_1 + q_2)^2 \frac{\epsilon^5}{Z^5} + ...
\end{equation}
Plugging in the cut-off in \eqref{CnmuF} we obtain the general result
\begin{equation}
{\cal{C}}^{(5)}[n, g \mu_F] = -\frac{27 \pi^2 n}{16 g^4 G_{N}^{(6)}} \left(\left(\frac{Z}{\epsilon }- \frac{q_1 + q_2}{8}\frac{\epsilon^2}{Z^2}\right)^3- y_{core}^3\right),
\end{equation}
from which it follows that the vacuum contribution is given by
\begin{equation}
{\cal{C}}^{(5)}[1,0] = -\frac{27 \pi^2 n}{16 g^4 G_{N}^{(6)}} \left(\frac{Z^3}{\epsilon^3}- 1\right).
\end{equation}
Finally, we arrive at
\begin{equation}\label{eq:cc6DmuF}
{\cal{C}}_D^{(5)} = {\cal{C}}^{(5)}[n,g \mu_F] - n {\cal{C}}^{(5)}[1,0] = \frac{10}{7} \left(\frac{1 +13 n +22 n^2 - g^2 \mu_F^2}{3^3 n^2} \sqrt{(1 + 2 n)^2- g^2 \mu_F^2} - 4 n \right) F^{\text{S}^5},
\end{equation}
where we have substituted $y_{core}$ as given by equation \eqref{eq:ycore6DU12} and $q_I$ as in \eqref{eq:qItomuF}. In turn, $F^{\text{S}^5}$ is the free energy of the 5d Sp(N) fixed point theory, computed in  \cite{Jafferis:2012iv} 
\begin{equation}\label{eq:ccAdS6bis}
F^{\text{S}^5} = \frac{9 \sqrt{2} \pi}{5}\frac{N^{5/2}}{ \sqrt{8-N_f}},
\end{equation}
which can be rewritten as
\begin{equation}\label{eq:ccAdS6}
F^{\text{S}^5} = \frac{3^5 \pi^4 \lambda ^4}{5 \, 2^8 g^8 G_N },
\end{equation}
using the flux quantisation conditions 
\begin{equation}
2 \pi F_0 = 8-N_f, \qquad \frac{1}{(2\pi)^3} \int_{I_{\xi} \times \text{S}^3} F_4 = N ,
\end{equation}
where $N_f$ stands for the number of D8-branes, and from which one obtains
\begin{equation}
8-N_f=\frac{2^2\pi g}{3\lambda^3}, \qquad N=\frac{3\lambda}{2^3 \pi g^3}.
\end{equation}

In \cite{Conti:2025wwf} (see also \cite{Arav:2024wyg,Arav:2024exg,Capuozzo:2023fll}) the defect entanglement entropies associated to co-dimension 2 monodromy defects in ABJM, $\mathcal{N}=4$ SYM and the 6d (2,0) CFT were related to a combination of the defect free energy/Weyl anomaly and the conformal weight. Similar results are not known however for such defects within 5d fixed point theories. In the current case, motivated by these other examples, we are encouraged to propose a relation of the form
\begin{equation}\label{eq:CDfreenergyconfweight5D}
{\cal{C}}_D^{(5)} = \alpha {\cal{I}}_D + \beta n h_D,
\end{equation}
with
\begin{equation}\label{eq:IDhD3d5D}
\begin{split}
{\cal{I}}_D & = \frac{40}{189 n^2 \alpha} \left(\left((1 + 2 n)^2 - g^2\mu_F^2\right)^{3/2}-27 n^3\right) F^{\text{S}^5}, \\[2mm]
h_D & = \frac{10 }{63 n^3 \beta} \left(-1-n +2 n^2+ g^2 \mu_F^2\right) \sqrt{(1 + 2 n)^2 - g^2 \mu_F^2} F^{\text{S}^5},
\end{split} 
\end{equation}
where $\alpha$ and $\beta$ are two constants independent of the monodromy sources. Our notation ${\cal{I}}_D, h_D$ suggests an analogy with the situation for other monodromy defects studied in the literature \cite{Arav:2024wyg,Arav:2024exg,Conti:2025wwf,ContiStuardo}. Indeed, one can notice that the quantity $h_D$ in \eqref{eq:IDhD3d5D} is proportional to $q_1+q_2$, as it happens for the conformal weights of the defects analysed in these references\footnote{In particular, the defect conformal weight is proportional to $q_1+q_2$ as a consequence of supersymmetry, which relates $h_D$ to the one-point function of the R-symmetry current $\langle J^R \rangle$. The latter is found to be proportional to $q_1+q_2$ \cite{Conti:2025wwf,Arav:2024wyg,Arav:2024exg} through a standard application of the holographic dictionary.}. In turn, ${\cal{I}}_D$ is proportional to $y_{core}(1,0)^3 - y_{core}(n,g\mu_F)^3$, which is the same behaviour found for the free energy/Weyl anomaly\footnote{More precisely, the free energy/Weyl anomaly is proportional to $y_{core}(1,0)^m - y_{core}(n,g\mu_F)^m$, where $m$ is a number that depends on the particular solution. In this case, $m=3$.}.
We would like to emphasise that despite the fact that in \eqref{eq:CDfreenergyconfweight5D} the quantities $\alpha$ and $\beta$ are numerical constants that cannot be fixed using our line of reasoning, we find the relations 
\eqref{eq:CDfreenergyconfweight5D}-\eqref{eq:IDhD3d5D} extremely suggestive, as they generalise known results in other dimensions. It would be very interesting to fix the constants $\alpha$ and $\beta$ following similar techniques to the ones applied in  \cite{Conti:2025wwf,Arav:2024exg,Arav:2024wyg} as well as to see whether a similar relation holds for more general 3d defects, as is the case in other dimensions.

\begin{figure}[t!]
 \centering
 	{{\includegraphics[width=12cm]{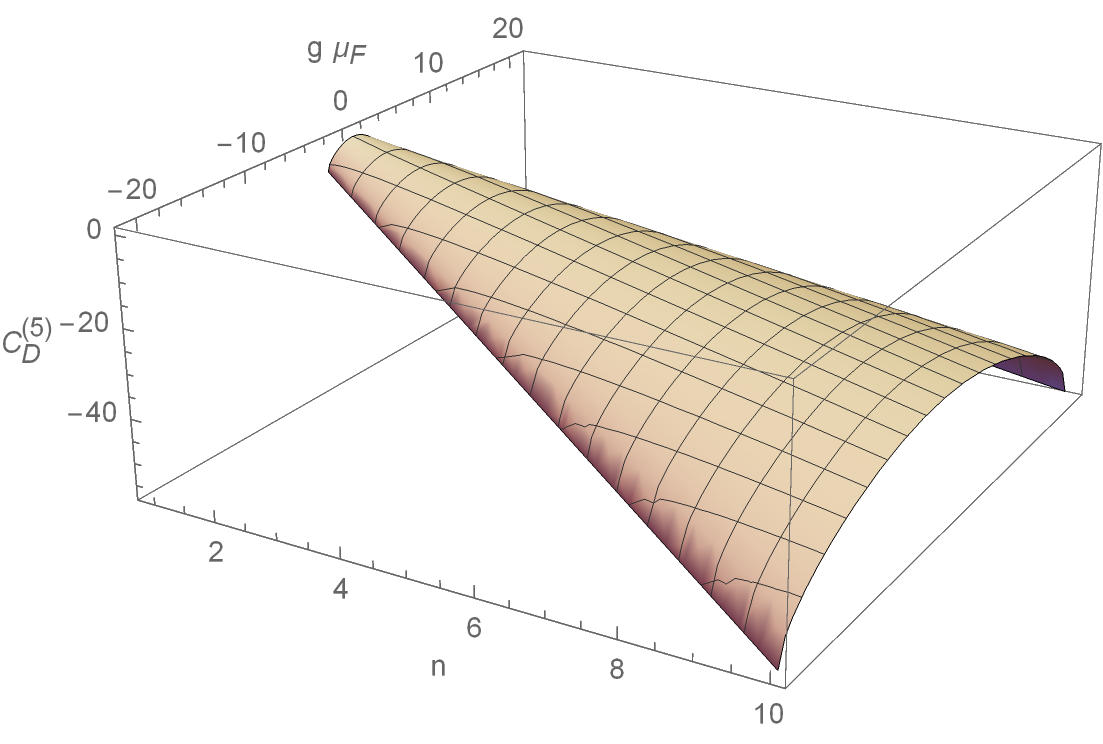} }}%
 	\caption{Plot of ${\cal{C}}_D^{(5)}$, in units of $F^{\text{S}^5}$, as a function of $n$ and $g \mu_F$.}
 	\label{Plot:6DccPlot}
 \end{figure}

We can analyse numerically the behaviour of ${\cal{C}}_D^{(5)}$ as a function of $n$ and $g \mu_F$, with $g \mu_F$ satisfying
\begin{equation}
-1 - 2n < g \mu_F < 1+2n,
\end{equation}
as implied by \eqref{eq:ycore6DU12}. The plot corresponding to such defects is depicted in Figure \ref{Plot:6DccPlot}.

Finally, note that in the two particular cases $q_2=0$ and $q_1=q_2$, $g \mu_F$ is further constrained to take specific values, given by \eqref{eq:muI6Dq2} and \eqref{eq:muI6Dminimal}, respectively. In these cases we find
\begin{equation}
q_2=0: \qquad \qquad {\cal{C}}_D^{(5)} = 10 \frac{\sqrt{3} \sqrt{2 + n} (5 + 7n) - 36 n^{3/2}}{63 \sqrt{n}} F^{\text{S}^5},
\end{equation}
and
\begin{align}
q_1=q_2: \qquad \qquad {\cal{C}}_D^{(5)} = 10 \frac{(1-n) (1 + 8 n)^2}{189 n^2} F^{\text{S}^5}.
\end{align}

\subsection{The $\text{AdS}_2\times S^1\times I\times \Sigma_k $ monodromy defect}\label{DefectinAdS4Sigma}

In this subsection we study a solution to 6d $U(1)^2$ matter-coupled gauged supergravity which upon uplift to massive Type IIA supergravity we interpret as describing surface defects within the 3d $\mathcal{N}=2$ CFTs dual to the class of $\text{AdS}_4$ solutions constructed in \cite{Bah:2018lyv}. Analogous solutions with the semi-compact space $S^1\times I$ replaced by a spindle were studied in \cite{Faedo:2021nub}, whose conventions we adopt. The solutions in \cite{Bah:2018lyv} are expected to be connected by RG flows to the $\text{AdS}_6$ vacuum of massive Type IIA supergravity \cite{Brandhuber:1998xy}, although to our knowledge such flows have not been constructed\footnote{See \cite{Hosseini:2018usu} for the BPS equations that these flows should satisfy in 6d $F(4)$ gauged supergravity.}. Similarly, it is natural to speculate that the solutions that we present in this subsection could be connected by RG flows to the $\text{AdS}_4\times S^1\times I$ solution to 6d $U(1)^2$ gauged supergravity discussed in subsection \ref{AdS4S1I6DU12}. The possible connections between these families of solutions are summarised in the following diagram.
\[
\begin{tikzcd}
 \text{AdS}_6 \times \text{S}^3 \times I \arrow[d, "Defect"'] \arrow[r, "RG flow"] &  \text{AdS}_4 \times \Sigma_k \times \tilde{\text{S}}^3 \times I \arrow[d, "Defect"] \\
\text{AdS}_4 \times \text{S}^1 \times I \times \tilde{\text{S}}^3 \times I  \arrow[r, "RG flow"'] & \text{AdS}_2 \times \text{S}^1 \times I \times \Sigma_k \times \tilde{\text{S}}^3 \times I
\end{tikzcd}
\]

The $\text{AdS}_2 \times \text{S}^1 \times I \times \Sigma_k $ solution we are interested in is obtained by uplifting to 6d $U(1)^2$ matter-coupled gauged supergravity the solution to 4d minimal supergravity discussed in \cite{Arav:2024wyg} (see also \cite{Ferrero:2020twa}), interpreted as describing surface monodromy defects in ABJM. The uplift rules for 6d $U(1)^2$ gauged supergravity coupled to a 2-form field were derived in \cite{Hosseini:2020wag}. Here we use the conventions for the Lagrangian and the consistent truncation in \cite{Faedo:2021nub}. The solution to 4d minimal gauged supergravity reads
\begin{equation}                              
\begin{split}\label{eq:4dminq1}
ds^2_4 & = \frac{h^2}{4} ds^2(\text{AdS}_2) +\frac{h^2}{P} d y^2+\frac{P}{4 h^2} n^2 d \varphi^2, \, \qquad A = \left(2 \frac{n y}{h} - 1-n\right) d \varphi, \\[2mm]
h  & = y + q, \qquad P  = h^4 - 4 y^2, \qquad   \varphi\in [0,2\pi],
\end{split}
\end{equation}
with $q$ a constant\footnote{Note that we have set the parameter $g=\frac{1}{2}$ compared to the solution discussed in \cite{Conti:2025wwf}. In this section, we will introduce a {\it new} parameter $g$ that will denote the coupling of the 6d theory.}.  $A$ has been chosen such that $A(y_{core}) = 0$. In this way
\begin{equation}
A(y \to \infty) = (n - 1) d \varphi.
\end{equation}
In turn, $y\in [y_{core},\infty)$ with $y_{core}$ taken as the highest root of the quartic polynomial $P$. Demanding that the $S^1\times I$ space spanned by $\varphi$ and $y$ closes smoothly at $y_{core}$ one finds
\begin{equation}\label{eq:ycore4Dminimal}
y_{core} = \frac{(1+n)^2}{2 n^2}, \qquad q = \frac{n^2-1}{2 n^2}.
\end{equation}

The uplift of this solution to 6d $U(1)^2$ gauged supergravity gives \cite{Hosseini:2020wag}
\begin{equation}\label{eq:AdS2S1ISigma}
\begin{split}
ds_6^2 & = e^{2A} \left[ \frac{h^2}{4} ds^2(\text{AdS}_2) + \frac{h^2}{P} \, dy^2 + \frac{P}{4 h^2} \, n^2 \, d\varphi^2 \right] + e^{2C} ds^2(\Sigma_k), \\[2mm]
F_1 & = dA_1= - \frac{q}{m} \frac{k_8^{1/2} k_2^{1/2}}{h^2} n \, d y \wedge d \varphi + \frac{k + {\mathtt{z}}}{2g} \text{vol}(\Sigma_k), \\[2mm]
F_2 & = dA_2= - \frac{q}{m}  \frac{k_8^{1/2} k_2^{-1/2}}{h^2} n \, d y \wedge d \varphi + \frac{k - {\mathtt{z}}}{2g} \text{vol}(\Sigma_k), \\[2mm]
X_1 & = k_8^{1/8} k_2^{1/2} ,  \qquad  X_2 = k_8^{1/8} k_2^{-1/2} , \qquad
{\cal B}_2 = - \frac{q}{m^2} k_8^{1/2} \, \text{vol}(\text{AdS}_2).
\end{split}
\end{equation}
Here $\Sigma_k$ is a 2d Riemann surface with constant curvature $k$\footnote{We will focus here on the $S^2$ ($k=1$) and $H^2$ ($k=-1$) cases.}. 
The warp factors are given by
\begin{equation}\label{AandC}
e^{2A} = \frac{k_8^{3/4}}{m^2}, \qquad e^{2C} = \frac{1}{m^2 k_8^{1/4} k_4},
\end{equation}
and the auxiliary parameters $k_2, k_4, k_8$ by
\begin{equation}
k_2 = \frac{ 3 {\mathtt{z}} + \sqrt{k^2 + 8 {\mathtt{z}}^2} }{ {\mathtt{z}} - k }, \qquad k_4  = \frac{18}{- 3 k + \sqrt{k^2 + 8 {\mathtt{z}}^2} }, \qquad k_8 = \frac{16 k_2}{9(1+k_2)^2}, 
\end{equation}
where $m = \frac{2 g}{3}$. Here ${\mathtt{z}}$, $k_2$, $k_8$ are parameters intrinsic to the AdS$_4 \times \Sigma_k $ solution (see \cite{Faedo:2021nub} for more details).
In particular, ${\mathtt{z}}$ is the twist parameter that ensures that the gauge fields are properly quantised when integrated over the Riemann surface\footnote{Setting ${\mathtt{z}}=0$ implies $k=-1$ and sets $A_1 = A_2$ and $X_1 = X_2$, rendering the resulting AdS$_4 \times H^2$ geometry a solution to 6d U($1$) gauged supergravity.}.

Compared to the solution constructed in \cite{Faedo:2021nub}, $y$ is now taken to live in the semi-compact interval $y\in [y_{core},\infty)$, with $y_{core}$ given by \eqref{eq:ycore4Dminimal}. 
One can check that setting $q=0$ the AdS$_4 \times \Sigma_k$ vacuum \cite{Karndumri:2015eta,Hosseini:2018usu} is restored, to which the solution asymptotes locally when $y\rightarrow\infty$,
\begin{equation}\label{eq:AdS4Sigma}
\begin{split}
ds_6^2 & = e^{2A} ds^2(\text{AdS}_4) + e^{2C} ds^2(\Sigma_k), \\[2mm]
F_1 &= \frac{k +  {\mathtt{z}}}{2g} \text{vol}(\Sigma_k), \qquad F_2 = \frac{k -  {\mathtt{z}}}{2g} \text{vol}(\Sigma_k), \\[2mm]
X_1 &= k_8^{1/8} k_2^{1/2}, \qquad  X_2 = k_8^{1/8} k_2^{-1/2},
\end{split}
\end{equation}
with $e^{2A}$ and $e^{2C}$ given by \eqref{AandC}. The asymptotic $\text{AdS}_4$ subspace contains again a conical singularity and there is a non-zero profile for the ${\cal B}_2$ field due to the monodromy.

The solution \eqref{eq:AdS2S1ISigma} can now be uplifted to massive Type IIA supergravity using the consistent truncation derived in \cite{Faedo:2021nub,Couzens:2022lvg}. The uplift formulas in \cite{Faedo:2021nub} allow one to extract the metric and dilaton of the uplifted solution. This was completed in \cite{Couzens:2022lvg} by  including the non-vanishing 2-form field of matter-coupled 6d $U(1)^2$ gauged supergravity, from which the background fluxes can be derived. Using the consistent truncation in \cite{Couzens:2022lvg} one finds the uplifted solution
\begin{equation}\label{10dexpression}
\begin{split}
ds_{10}^2 & = \frac{\lambda^2}{\mu_0^{1/3}} \bigg[ \tilde{\Delta}^{1/2} \, ds_6^2 + \frac{1}{g^2 \tilde{\Delta}^{1/2} k_8^{1/4}} \bigg( k_8^{1/2} d \mu_0^2+ \frac{1}{\sqrt{k_2}} \left( d \mu_1^2 + \mu_1^2 (d \phi_1 - g A_1)^2 \right) \\[2mm]
& + \sqrt{k_2} \left( d \mu_2^2 + \mu_2^2 (d \phi_2 - g A_2)^2 \right) \bigg) \bigg], \\[2mm]
e^{\Phi} & = \frac{\lambda^2 \tilde{\Delta}^{\frac{1}{4}} }{\mu_0^{5/6} k_8^{1/8}}, \qquad B_2= \frac{\lambda^2}{2} \mu_0^{2/3} \frac{9 q}{4 g^2} k_8^{1/2} \, \text{vol}(\text{AdS}_2), \qquad F_2= \frac{1}{\lambda} \mu_0^{2/3} \frac{3 q}{4 g} k_8^{1/2} \, \text{vol}(\text{AdS}_2)
\end{split}
\end{equation}
where
\begin{equation}
\tilde{\Delta} =  \frac{1}{\sqrt{k_8}} \mu_0^2 + \sqrt{k_2} \mu_1^2 + \frac{1}{\sqrt{k_2}} \mu_2^2 
\end{equation}
and we have omitted the unwieldy expression for the $F_4$ flux (that can in any case be obtained from the uplift formulas in \cite{Couzens:2022lvg}).   
One can check that the metric and dilaton asymptote locally to the metric and dilaton of the uplift of the AdS$_4 \times \Sigma_k$ vacuum \eqref{eq:AdS4Sigma}, which, as shown in \cite{Faedo:2021nub}, is the $\text{AdS}_4\times \Sigma_k$ solution to massive Type IIA supergravity constructed in \cite{Bah:2018lyv}. As in previous examples, the $\text{AdS}_4$ subspace is deformed by the non-trivial monodromy sources and there are extra fluxes proportional to them. We will interpret the solution \eqref{10dexpression} as describing line defects within a twisted compactification of the 5d Sp(N) fixed point theory on $\Sigma_k$.

\subsubsection{Defect entanglement entropy}\label{holocc6d}

As in previous sections the entanglement entropy of the AdS$_2 \times S^1 \times I \times \Sigma_k$ solution can be computed using the prescription \eqref{eq:Ceq}-\eqref{eq:backsub}. We first find
\begin{equation}
{\cal{C}}^{(2)}[n] = \frac{n \pi}{4 G_N^{(2)}} ( \Lambda - y_{core} ).
\end{equation}
The cut-off in this case is given by
\begin{equation}\label{cut-off6dSigma}
\Lambda = \frac{Z}{\epsilon} - q - \frac{1}{2} q^2 \frac{\epsilon}{Z}+\dots
\end{equation}
such that
\begin{equation}
{\cal{C}}^{(2)}[n] = \frac{n \pi}{4 G_N^{(2)}} \left( \frac{Z}{\epsilon} - q - y_{core} \right).
\end{equation}
Finally we can compute the defect entanglement entropy using the prescription $ {\cal{C}}_D^{(2)} = {\cal{C}}^{(2)}[n] - n {\cal{C}}^{(2)}[1]$. This gives the defect contribution
\begin{equation}
{\cal{C}}_D^{(2)} = \frac{\pi}{4} (n-1) F^{\text{S}^3 \times \Sigma_k} ,
\end{equation}
where $y_{core}$ and $q$ have been substituted by their expressions in \eqref{eq:ycore4Dminimal}, and $F^{\text{S}^3 \times \Sigma_k} $ stands for the free energy of the 3d CFT that arises in the IR upon compactifying the 5d Sp(N) fixed point theory, given in our conventions by
\begin{equation}
F^{\text{S}^3 \times \Sigma_k} = \frac{2^{3/2}}{9 \pi} \frac{(\mathtt{z}^2 - k^2)^{3/2}(\sqrt{k^2 + 8 \mathtt{z}^2} - k ) }{ k \sqrt{k^2 + 8 \mathtt{z}^2} + 4 \mathtt{z}^2 - k^2} \text{Vol}(\Sigma_k) F^{\text{S}^5} ,
\end{equation}
with  $F^{\text{S}^5}$ as in \eqref{eq:ccAdS6bis}. We note that we reproduce the same $n$ dependence found in \cite{Conti:2025wwf}. 

\subsection{The $\text{AdS}_2\times S^1\times I\times {\text{spindle}}$ monodromy defect}\label{defectspindle}

In this subsection we study a solution to 6d $U(1)^2$ gauged supergravity that, once uplifted to massive Type IIA supergravity, can be interpreted as describing co-dimension 2 defects within the 5d Sp(N) fixed point theory compactified on a spindle  \cite{Faedo:2021nub}. This solution is locally equivalent to the solutions constructed in \cite{Couzens:2022lvg,Faedo:2022rqx}. However, globally one of the spindles is replaced by a semi-compact space $S^1\times I$. 

The solution we are interested in is obtained uplifting the solution of 4d minimal supergravity studied in \cite{Arav:2024wyg}, and summarised in  \eqref{eq:4dminq1}-\eqref{eq:ycore4Dminimal}, to 6d $U(1)^2$ matter-coupled gauged supergravity, but now using the consistent truncation constructed in \cite{Couzens:2022lvg}. 
We obtain
\begin{eqnarray}\label{eq:AdS26D}
&&ds_6^2 = \frac{ 3^{3/2} \left(y^2 h_1 h_2 \right)^{1/4}}{2^{3/2} g^{3/2} m^{1/2}}  \left[ \frac{h_4^2}{4} ds^2(\text{AdS}_2) + \frac{h_4^2}{P_4} dx^2 + \frac{P_4}{4 h_4^2} n^2 d \varphi^2  + \frac{y^2}{P} dy^2 + \frac{P}{h_1 h_2 } \left(d\phi - \frac{1}{2} A \right)^2\right]\,, \nonumber\\
&&A_I = \alpha_I d \phi - \frac{3}{2 g} \frac{y^3}{h_I} \left(d\phi-\frac{1}{2} A \right), \quad X_I = \frac{3^{1/3} m^{1/4}}{2^{1/4} g^{1/4}}\frac{\left(y^2 h_1 h_2 \right)^{3/8}}{h_I}, \quad {\cal B}_2 = - \frac{3}{2 g m}y*_4 F,\nonumber\\
\end{eqnarray}
with $A$ the 4d gauge field,
\begin{equation}
A = \Bigl(\frac{2 nx}{h_4}-n-1\Bigr) d\varphi
\end{equation}
and $F$ its field strength. Note that we have renamed
\begin{equation}\label{changenot}
h\rightarrow h_4,\quad P\rightarrow P_4,\quad y\rightarrow x
\end{equation}
as compared to \eqref{eq:4dminq1}-\eqref{eq:ycore4Dminimal}, and introduced 
\begin{equation}
h_I = y^3 + q_I, \quad I=1,2, \qquad P  = h_1 h_2 - y^4.
\end{equation}
In this solution we take $\varphi$ and $x$ to span the $S^1\times I$ semi-compact space of the solution to 4d minimal gauged supergravity studied in \cite{Arav:2024wyg} and $y$ and $\phi$ to span a spindle\footnote{The $\alpha_I$ in \eqref{eq:AdS26D} are fixed so as to ensure that the spindle is well-defined.}. deformed by the monodromies. By construction this solution asymptotes locally (when $x$ approaches infinity) to the $\text{AdS}_4\times \spindle$ solution to 6d $U(1)^2$ gauged supergravity constructed in \cite{Faedo:2021nub}, with the spindle, $\spindle$, parametrised by $y$ and $\phi$. 
The uplift of the solution  to massive Type IIA supergravity can then be interpreted as describing co-dimension 2 defects within the 5d Sp(N) fixed point theory compactified on a spindle, the dual of which was constructed in \cite{Faedo:2021nub}.

The solution in \cite{Faedo:2021nub} is expected to be connected by an RG flow to the Brandhuber-Oz solution\footnote{See \cite{Ferrero:2020twa} for black hole solutions interpolating between $\text{AdS}_2 \times \text{spindle}$ and $\text{AdS}_4$ in 4d minimal gauged supergravity.}. Therefore, one may speculate that the uplift to massive Type IIA of the $\text{AdS}_2$ solution \eqref{eq:AdS26D}  could be connected by an RG flow to the  $\text{AdS}_4 \times \text{S}^1 \times I \times \text{S}^3 \times I$ solution presented in subsection 
\ref{6d-uplift-to10d}. These conjectural connections are depicted in the following diagram, where $\spindle$ represents the spindle.
\[
\begin{tikzcd}
 \text{AdS}_6 \times \text{S}^3 \times I \arrow[r, "RG flow"] \arrow[d, "Defect"'] & \text{AdS}_4 \times \spindle \times \tilde{\text{S}}^3 \times I \arrow[d, "Defect"] \\
\text{AdS}_4 \times \text{S}^1 \times I \times \tilde{\text{S}}^3 \times I 
\arrow[r, "RG flow"'] & \text{AdS}_2 \times \text{S}^1 \times I \times \spindle \times \tilde{\text{S}}^3 \times I
\end{tikzcd}
\]

Finally, we can compute the defect entanglement entropy of the uplifted solution, to find
\begin{equation}
\begin{split}
{\cal{C}}_D & = \frac{4 \pi n}{81} (2- x_{core} - q) (y_S^3 - y_N^3) F^{\text{S}^5}, \\[2mm]
& = \frac{\pi}{4} (n-1) F^{\text{S}^3 \times \spindle} ,
\end{split}
\end{equation}
where $x_{core}$ and $q$ have been substituted using \eqref{eq:ycore4Dminimal}\footnote{Note the different conventions.}. In this expression $y_S$ and $y_N$ are the two roots of $P_4$ associated to the south and north poles of the spindle. Their explicit expressions can be found in \cite{Faedo:2021nub}. Indeed
\begin{equation}
F^{\text{S}^3 \times \spindle} = \frac{16}{81} (y_S^3 - y_N^3) F^{\text{S}^5}
\end{equation}
is the free energy associated to the 3d CFT that arises in the IR upon compactifying the 5d Sp(N) fixed point theory on a spindle \cite{Faedo:2021nub}. Instead, the $(n-1)$ factor in front is the contribution from the defects (up to normalisation), and agrees with the expression found in \cite{Conti:2025wwf} in the limit in which the four gauge fields of the 4d $U(1)^4$ theory considered therein coincide. 

Finally, we would like to point out that one could also consider taking the coordinates $\phi$ and $y$ to span a second semi-compact space $S^1\times I$ instead of a spindle. The resulting solution asymptotes locally to the $\text{AdS}_6$ vacuum when $x$ and $y$ are sent to infinity, suggesting that a defect interpretation might be possible.

\section{Conclusions}\label{conclusions}

In this work we have studied several solutions to massive Type IIA supergravity that we have interpreted as describing monodromy defects within 6d (1,0) CFTs, the 5d Sp(N) fixed point theory or compactifications thereof. 

We began our analysis with the study of a solution to 7d $U(1)$ gauged supergravity whose metric can be written as a foliation by $\text{AdS}_5\times S^1$ over an interval, and asymptotes locally to the maximally supersymmetric $\text{AdS}_7$ vacuum. The uplift of this solution to massive Type IIA supergravity gives rise to a class of solutions that asymptote locally to the $\text{AdS}_7$ solutions in \cite{Apruzzi:2013yva}, and can be interpreted as describing 4d defects within the 6d (1,0) CFTs dual to this class of solutions. Multi-charged  solutions to 7d $U(1)^2$ gauged supergravity can also be constructed, but they do not admit an uplift to massive Type IIA supergravity, which is our common thread in this work. For this reason we have relegated them to an Appendix.  The analysis of defects within the 5d Sp(N) fixed point theory starts with the study of solutions to 6d $U(1)^2$ gauged supergravity that are foliations of the six-dimensional spacetime by $\text{AdS}_4\times S^1$ on an interval and approach the maximally supersymmetric $\text{AdS}_6$ vacuum asymptotically. The uplift to massive Type IIA supergravity gives rise to a solution that asymptotes locally to the Brandhuber-Oz solution, and can be interpreted as describing 3d defects within the 5d Sp(N) fixed point theory. 

As we have stressed throughout the paper, our solutions are locally equivalent to the spindle solutions constructed in \cite{Ferrero:2021wvk,Faedo:2021nub}. Therefore, they can also be obtained upon double analytical continuation \cite{Lu:2003iv} from the black hole solutions found in \cite{Cvetic:1999ne,Cvetic:1999xp,Liu:1999ai}. An important difference is that globally the spindles are replaced by semi-compact spaces $S^1\times I$, making possible the defect interpretation. Similar constructions of co-dimension 2 holographic defects have appeared in \cite{Gutperle:2022pgw,Gutperle:2023yrd,Arav:2024exg,Arav:2024wyg}. As in those works, the defects studied in this paper are interpreted as monodromy defects, because they have associated background gauge fields that provide non-trivial monodromy sources for charged matter which circumnavigates the defect.

We have completed our analysis with the construction of surface monodromy defects or line monodromy defects within the  6d or 5d theories compactified on a Riemann surface with constant curvature or a spindle. In the first case we have treated in the main text the compactification of the 6d theories on a hyperbolic plane and relegated to an Appendix the compactifications that cannot be uplifted to massive Type IIA supergravity. These solutions are locally equivalent to solutions constructed in the literature describing compactifications of the 6d or 5d theories on a spindle and a Riemann surface or on two spindles \cite{Boido:2021szx,Faedo:2021nub,Giri:2021xta,Cheung:2022ilc,Suh:2022olh,Couzens:2022lvg,Faedo:2022rqx,Hristov:2024qiy}. Globally the spindle (or one of the spindles in the second case) is replaced by a $S^1\times I$ semi-compact space, making possible the defect interpretation. Replacing the two spindles by two semi-compact spaces $S^1\times I$ could be a possible way to describe co-dimension 4 monodromy defects or monodromy defects within defects, that should be interesting to explore.

All the solutions discussed in this paper are accompanied by a detailed calculation of the contribution of the defect to the entanglement entropy, complementing our findings in \cite{Conti:2025wwf} for co-dimension 2 monodromy defects. In particular, we have seen that for 3d monodromy defects embedded in the 5d Sp(N) fixed point theory it is possible to find an expression that relates the defect entanglement entropy to a linear combination of a quantity which we associate (by analogy) to  the defect free energy and the conformal weight of the defect, mimicking previous results for 2d  \cite{Jensen:2018rxu,Kobayashi:2018lil}, 4d  \cite{Chalabi:2021jud} and 1d conformal defects. In the last case, relevant for defects within the ABJM theory, it was found in \cite{Conti:2025wwf} that the ``defect free energy'' plays the role of the Weyl anomaly for analogous even dimensional cases. It would be very interesting to further investigate whether this relation extends to broader classes of defects beyond the monodromy ones studied in this paper. 

Furthermore, it would be interesting to see whether the relation \eqref{eq:CDfreenergyconfweight5D} is also satisfied in non-supersymmetric settings, as it is the case for the 2d and 4d defects studied in  \cite{Jensen:2018rxu,Kobayashi:2018lil,Chalabi:2021jud}. The solutions discussed in this paper can be made non-supersymmetric by switching on the $\nu$-parameter that was set to zero in equation \eqref{eq:6DsolutionU12}. It should be simple to check whether equation \eqref{eq:CDfreenergyconfweight5D} still holds for these solutions. Finally, it would be interesting to provide explicit prescriptions for the calculation of the defect entanglement entropy for more general classes of defects.  In particular it would be interesting to confirm whether the $\text{AdS}_3$ solutions constructed in \cite{Lozano:2022ouq} should be interpreted as duals to deconstructed 6d (1,0) CFTs, as claimed in \cite{Conti:2024qgx}, rather than to co-dimension 4 defects.

We speculate on the existence of RG flows relating the defect CFTs in which the ambient theories are compactifications of the 6d (1,0) theories or the 5d Sp(N) fixed point theory (both on a Riemann surface and on a spindle), after flowing to the IR, and the UV defect CFTs whose ambient theories are the respective (uncompactified) CFTs. It would be very interesting (though likely technically challenging) to construct the explicit solutions holographically realising these flows. This would provide an excellent arena to search for and evaluate proposed RG monotones in the defect setting. Previous results in the literature demonstrate that such monotonicity can be elusive  \cite{Herzog:2019rke,Conti:2024qgx}. 

Defect CFTs are typically realised holographically in terms of brane intersections that involve both defect and ambient branes. This description has been provided in many constructions of holographic defects found in the literature. See for instance \cite{Dibitetto:2017tve,Dibitetto:2017klx,Dibitetto:2018iar,Dibitetto:2018gtk,Faedo:2020nol,Faedo:2020lyw,Lozano:2021fkk,Lozano:2022vsv,Lozano:2022swp,Lozano:2022ouq}. In this paper we have proposed one such brane intersection for the co-dimension 2 monodromy defects embedded in the 6d (1,0) CFTs dual to the $\text{AdS}_7$ solutions in  \cite{Apruzzi:2013yva}. This brane intersection has been proposed based on the analysis of the background fluxes present in the defect solution, namely, after the near-horizon limit has been taken. It would be interesting to identify the brane intersection prior to taking the near horizon limit. Pursuing further these investigations may allow us to further refine the AdS/CFT correspondence for defect CFTs, by comparing the defect entanglement entropies (or central charges/free energies) with the corresponding field theory observables. This would be aligned with the investigations of surface defects in 6d (1,0) CFTs carried out in \cite{Faedo:2020nol,Lozano:2022ouq,Conti:2024qgx}, whose holographic central charges were reproduced from field theory calculations. 

In Type IIB, 3d defects in 5d CFTs with $\text{AdS}_6\times S^2\times \Sigma_2$ geometric duals have been studied in the probe brane approximation, adding D3-branes to the 5-brane webs underlying the 5d CFTs  \cite{Santilli:2023fuh,Gutperle:2020rty}. It would be interesting to investigate whether the uplift to Type IIB supergravity of the 3d monodromy defects considered in this paper bears any relation with these constructions, once the backreaction of the defects has been taken into account.

Finally, all  the monodromy defects studied in this paper preserve conformal symmetry. Similar techniques have been applied to supersymmetric non-conformal monodromy defects in \cite{ContiStuardo}.

\section*{Acknowledgements}
We would like to thank Chris Couzens, Jerome Gauntlett, Niall Macpherson, Carlos Nunez, Achilleas Passias and Alessandro Tomasiello for useful discussions.
AC and YL are partially supported by the grants from the Spanish government MCIU-22-PID2021-123021NB-I00 and MCIU-25-PID2024-161500NB-I00. The work of AC is also supported by the Severo Ochoa fellowship PA-23-BP22-019. AC acknowledges the INGENIUM Alliance of European Universities for giving the opportunity to spend training time at the University of Crete. AC thanks Crete Center for Theoretical Physics and Imperial College London for the kind hospitality while some parts of this work were being completed. The work of CR is supported through the framework of H.F.R.I. call ``Basic research Financing (Horizontal support of all Sciences)'' under the National Recovery and Resilience Plan ``Greece 2.0'' funded by the European Union--NextGenerationEU (H.F.R.I. Project Number: 15384). Y.L. would like to thank the Isaac Newton Institute for Mathematical Sciences, Cambridge, for support and hospitality during the programme "Boundaries, Impurities and Defects" where work on this paper was undertaken. This work was supported by EPSRC grant no EP/R014604/1.$\& \#$34.

\appendix

\section{Monodromy defects in M-theory}\label{appendix}

In this Appendix we present several extensions of our analysis in the main body of the paper that do not admit a description in massive Type IIA supergravity, but can be interpreted within M-theory. One obvious extension is the generalisation of the $\text{AdS}_5\times S^1\times I$ monodromy defects constructed in section \ref{6dtheories} to defects involving one extra monodromy parameter, realised as solutions to 7d $U(1)^2$ gauged supergravity. These solutions do not admit an uplift to massive Type supergravity, but can be uplifted to M-theory, where they asymptote locally to the maximally symmetric $\text{AdS}_7\times S^4$ vacuum. They can  then be interpreted as describing co-dimension 2 defects within the 6d (2,0) CFT. These solutions were studied in \cite{Conti:2025wwf}, so we omit them in this Appendix, and focus on $\text{AdS}_3$ solutions to 7d $U(1)^2$ gauged supergravity  involving constant curvature Riemann surfaces and spindles and their uplifts to M-theory. These solutions describe co-dimension 2 monodromy defects within compactifications of the 6d (2,0) CFT on a Riemann surface with constant curvature or a spindle, after flowing to the IR. 

\subsection{The $\text{AdS}_3\times S^1\times I\times \Sigma_k$ monodromy defect}\label{AdS3Riemann}

In this sub-appendix we extend the analysis in subsection \ref{AdS3H2section} to the case in which the $\text{AdS}_3\times S^1\times I\times \Sigma_k$ geometry is a solution to 7d $U(1)^2$ gauged supergravity. This allows us to consider a two dimensional Riemann surface with arbitrary constant curvature, $\Sigma_k$. The M-theory uplift of this solution asymptotes locally to the  AdS$_5 \times \Sigma_k\times {\tilde S}^4$ solutions constructed in \cite{Bah:2012dg}, and can thus be interpreted as describing co-dimension 2 monodromy defects within the 4d $\mathcal{N}=1$ CFTs dual to this class of solutions. Given that these solutions are connected by RG flows to the $\text{AdS}_7\times S^4$ M-theory vacuum \cite{Bah:2012dg}, one could expect that the uplift of the $\text{AdS}_3\times S^1\times I\times \Sigma_k$ solution could be connected by an RG flow to the $\text{AdS}_5\times S^1\times I$ solution studied in \cite{Conti:2025wwf}. These possible connections are depicted in the following diagram.
\[
\begin{tikzcd}
 \text{AdS}_7 \times \text{S}^4 \arrow[d, "Defect"'] \arrow[r, "RG flow"]   & \text{AdS}_5 \times \Sigma_{k} \times \tilde{\text{S}}^4 \arrow[d, "Defect"] \\
 \text{AdS}_5 \times \text{S}^1 \times I \times \tilde{\text{S}}^4 \arrow[r, "RG flow"'] &
 \text{AdS}_3 \times \text{S}^1 \times I \times \Sigma_{k} \times \tilde{\text{S}}^4 
 \end{tikzcd}
\]

The solution to 7d $U(1)^2$ gauged supergravity we are interested in is locally equivalent to the $\text{AdS}_3 \times \text{spindle} \times \Sigma_k$ solution constructed in \cite{Boido:2021szx}. However, globally the spindle is replaced by a semi-compact space $S^1\times I$. The solution reads
\begin{equation}\label{AdS3U(1)2}
\begin{split}
ds^2_7 & = e^{2 A}  \left( h ds^2(\text{AdS}_3) + \frac{h}{4 P} dy^2 + \frac{P}{h^2} d\varphi^2 \right) + e^{2C} ds^2(\Sigma_k), \\[2mm]
F_1 & = - \frac{2}{g} e^{A} X_1 \, k \, d A + \frac{ 1 + \mathtt{z}}{ 2 g} \text{vol}(\Sigma_k), \qquad F_2 = - \frac{2}{g} e^{A} X_2 \, k \,d A + \frac{1 - \mathtt{z}}{ 2 g} \text{vol}(\Sigma_k), \\[2mm]
{\cal{S}}^5 & = \frac{8}{g^2} e^{2 A} X_1 X_2 (h-y) \text{vol}(\text{AdS}_3), \qquad
X_1 = e^{2 \lambda_1}, \qquad X_2 = e^{2 \lambda_2},
\end{split}
\end{equation}
where $h$, $P$ and $A$ are as defined in \eqref{eq:5Dmin}, \eqref{handP}.
The warp factors are
\begin{equation}
e^A = X_1^2 X_2^2, \qquad e^C = \sqrt{X_1} \sqrt{X_2} \sqrt{- \frac{k}{8} \left( (1-\mathtt{z}) X_1 + (1+\mathtt{z}) X_2 \right)}
\end{equation}
and the scalars
\begin{equation}
\begin{split}
e^{10\lambda_1} & = \frac{1+7\,\mathtt{z}+7 \,\mathtt{z}^2 + 33\,\mathtt{z}^3 + k (1+4\,\mathtt{z}+19\,\mathtt{z}^2)\sqrt{1+3\,\mathtt{z}^2}}{4\,\mathtt{z}(1-\mathtt{z})^2} , \nn \\
e^{2(\lambda_1-\lambda_2)} & = \frac{1+\mathtt{z}}{2\,\mathtt{z}- k \sqrt{1+3\,\mathtt{z}^2}}.
\end{split}
\end{equation}
In turn, $y\in [y_{core},\infty)$ with $y_{core}$ given by \eqref{ycoreandq} and $k$ is the curvature of the Riemann surface.

This solution is very similar to the solution presented in subsection \ref{DefectinAdS4Sigma}. Also here $\mathtt{z}$ is the twist parameter, that ensures that the charges associated to the background fluxes are properly quantised \cite{Bah:2012dg}. The case $\mathtt{z}=0$ implies $k=-1$ (i.e. that the Riemann surface is the hyperbolic plane) and $A_1 = A_2$, $X_1 = X_2$. The solution then reduces to the one discussed in subsection \ref{AdS3H2section}, that can be uplifted to massive Type IIA supergravity. Instead, for $\mathtt{z}\neq 0$ the solution \eqref{AdS3U(1)2} can be uplifted to M-theory using the formulas derived in \cite{Cheung:2022ilc}. One gets
\begin{align}
ds^2_{10} & = \frac{4 \Delta^{1/3} e^{2 A}}{g^2}  \left(h ds^2(\text{AdS}_3) + \frac{h}{4 P} dy^2 + \frac{P}{h^2} d\varphi^2 \right) + \frac{4 \Delta^{1/3} e^{2C}}{g^2}  ds^2(\Sigma_k) \nn \\[2mm]
& + \frac{\Delta^{-2/3}}{g^2}\left( \frac{1}{X_0} d \mu_0^2 + \frac{1}{X_1} (d \mu_1^2 + \mu_1^2 (d \phi_1 - g A_1)^2 ) + \frac{1}{X_2} (d \mu_2^2 + \mu_2^2 (d \phi_2 - g A_2)^2 )\right),
\end{align}
where
\begin{equation} \label{eq:DeltaX05d7d11d}
\Delta = \mu_0^2 X_0 + \mu_1^2 X_1 + \mu_2^2 X_2, \qquad X_0 = \frac{1}{X_1^2 X_2^2},
\end{equation}
and $\{\mu_0,\mu_1,\mu_2\}$ satisfy $\sum_{a=0}^2 \mu_a^2=1$. It can be checked that this asymptotes locally to the metric of the AdS$_5 \times \Sigma_k\times {\tilde S}^4$ solution constructed in \cite{Bah:2012dg}. We have just presented the metric of the uplifted solution because in the reminder of this subsection we will focus on the computation of the defect entanglement entropy\footnote{For the expression of the fluxes we refer to \cite{Cheung:2022ilc}.}. Using our prescription in \eqref{eq:Ceq}-\eqref{eq:backsub} we find
\begin{equation}
{\cal{C}}_D^{(4)} = \frac{\pi^3}{108 n} (n-1) (5 n+1) \text{Vol}(\Sigma_k) e^{3A} e^{2 C} a^{(2,0)},
\end{equation}
where in our conventions
\begin{equation}
a^{(2,0)} = \frac{4 N^3}{\pi^3}, \qquad g = \frac{1}{\pi^{1/3} N^{1/3} }.
\end{equation}
Notice that we find the same defect + compactification contribution found for the $\text{AdS}_3\times S^1\times I\times H^2$ solutions studied in subsection \ref{AdS3H2section}, with the compactification piece, given by $e^{3A} e^{2 C} \text{Vol}(\Sigma_k)$, in agreement with the results in \cite{Bah:2012dg}.
Inserting the definition of the warp factors we finally obtain
\begin{equation}
{\cal{C}}_D^{(4)} = \frac{\pi^3}{18 n} (n-1)(1 + 5n) \frac{k - 9 k \mathtt{z}^2 + \left( 1 + 3 \mathtt{z}^2\right)^{3/2}}{769 \mathtt{z}^2} \text{Vol}(\Sigma_k) a^{(2,0)}.
\end{equation}

\subsection{The $\text{AdS}_3\times S^1\times I\times \text{spindle}$ monodromy defect}

In this sub-appendix we present a very similar construction to the one discussed in subsection \ref{defectspindle}, in this case for $\text{AdS}_3$. Our starting point is the double spindle solution to 7d $U(1)^2$ gauged supergravity constructed in \cite{Cheung:2022ilc}, where we replace one of the spindles with a semi-compact space $S^1\times I$. The solution is then uplifted to M-theory, where we interpret it as describing surface defects within the 6d (2,0) theory compactified on a spindle, described holographically by the $ \text{AdS}_5 \times \spindle \times \tilde{\text{S}}^4 $ solution obtained in \cite{Ferrero:2021wvk}. The M-theory uplift of the solution asymptotes locally to this background. The solution in \cite{Ferrero:2021wvk} is expected to be connected by an RG flow to the $\text{AdS}_7\times S^4$ vacuum. Analogously, one could expect that our solution in this sub-appendix could be connected by an RG flow to the uplift of the $\text{AdS}_5\times S^1\times I$ solution discussed in \cite{Conti:2025wwf}. These possible connections are depicted in the following diagram.
\[
\begin{tikzcd}
 \text{AdS}_7 \times \text{S}^4    \arrow[d, "Defect"']  \arrow[r, "RG flow"] & \text{AdS}_5 \times \spindle \times \tilde{\text{S}}^4 \arrow[d, "Defect"] \\
 \text{AdS}_5 \times \text{S}^1 \times I \times \tilde{\text{S}}^4  \arrow[r, "RG flow"'] & \text{AdS}_3 \times \text{S}^1 \times I  \times \spindle \times \tilde{\text{S}}^4
 \end{tikzcd}
\]

The solution we are interested in is the 7d analogue of the 6d solution \eqref{eq:AdS26D}, namely\footnote{Where in the conventions of \cite{Cheung:2022ilc} $A_1 = A_{12}$ and $A_2 = A_{34}$.}
\begin{eqnarray} 
ds^2_7 & = &\frac{(yh_1h_2)^{1/5}}{g^2} \left[ h_5 ds^2(\text{AdS}_3) + \frac{h_5}{4 P_5} dx^2 + \frac{P_5}{ h_5^2} n^2 d \varphi^2 +\frac{y}{P}d y^2+\frac{P}{h_1h_2} \left( d \phi - A \right)^2 \right] , \nn \\[2mm]
A_I& = &\alpha_I d \phi + \frac{2}{g}  \frac{y^2}{h_I} \left(d \phi - A \right) , \qquad 
A = \left( \alpha_0 + n \, \frac{x}{h_5} \right) d\varphi, \qquad F=dA, \nn \\[2mm]
X^I & = &\frac{(yh_1h_2)^{2/5}}{h_I}, \qquad
{\cal{S}}_3  = - \frac{1}{g^2} y \star_4 F + \frac{1}{g^2} \frac{y P}{h_1 h_2} \left(d \phi - A  \right) \wedge F, \label{eq:AdS37D}
\end{eqnarray}
where we have renamed
\begin{equation}\label{redefinitions}
h \to h_5, \qquad P \to P_5, \qquad y\to x
\end{equation}
as compared to \eqref{eq:5Dmin}, and introduced
\begin{equation}
h_I  = y^2+q_I ,\qquad I=1,2 , \qquad P = h_1h_2-4 y^3.
\end{equation}
As in  \eqref{eq:AdS26D} we take $\varphi$ and $x$ to span a $S^1\times I$ semi-compact space and $y$ and $\phi$ to span a spindle, with $\alpha_I$ defined accordingly. \footnote{Similar to \ref{defectspindle}, also for \eqref{eq:AdS37D} one could take the coordinate $y$ to be semi-infinite, such that \eqref{eq:AdS37D} would asymptote locally to the maximally supersymmetric $\text{AdS}_7$ vacuum when $x$ and $y$ are sent to infinity. This analysis would require a more detailed investigation, that we hope to report in a future publication.}

The previous solution can be uplifted to M-theory using the consistent truncation constructed in \cite{Cheung:2022ilc}. We obtain
\begin{align}
ds_{11}^2 & = \Delta^{1/3} \frac{(yh_1h_2)^{1/5}}{g^2} \left( h_5 ds^2(\text{AdS}_3) + \frac{h_5}{4 P_5} dx^2 + \frac{P_5}{ h_5^2} n^2 d \varphi^2 +\frac{y}{P}d y^2+\frac{P}{h_1h_2} \left(d \phi - A \right)^2 \right) \nn \\[2mm]
& + g^{-2} \Delta^{-2/3} \left( X_0^{-1} d\mu_0^2 + \sum_{I=1}^2 (X^I)^{-2} (d\mu_I^2 + \mu_I^2 (d\phi_I + g A^I )) \right),
\end{align}
where again we are just presenting the metric, and we have the same definitions for $\mu_a$, $a=0,1,2$, $\Delta$ and $X_0$ as in the previous subsection.

The defect entanglement entropy of the uplifted solution reads
\begin{align}
{\cal{C}}_D^{(4)} & = \frac{n \pi^4}{3 2^8} \left( 1- x_{core} - q_5 \right) (y_S^2 - y_N ^2) a^{(2,0)}, \nn \\[2mm]
& = \frac{\pi}{2} \frac{(n-1)(1 + 5 n)}{n} a^{(2,0) \times \spindle},
\end{align}
where we used the definitions \eqref{ycoreandq} and the cut-off \eqref{eq:cutoff5dminimal}\footnote{Note here our different notation \eqref{redefinitions}.}. $y_S$ and $y_N$ are the two roots of $P$ that define the spindle. In particular,
\begin{equation}
a^{(2,0) \times \spindle} = \frac{\pi^3}{384} (y_S^2 - y_N^2) a^{(2,0)}
\end{equation}
provides the ``compactification contribution" to the entanglement entropy, and matches the result in \cite{Ferrero:2021etw,Ferrero:2021wvk} (for more details we refer to \cite{Ferrero:2021etw,Ferrero:2021wvk,Cheung:2022ilc}). The contribution of the defects is encoded in the pre-factor $\frac{(n-1)(1 + 5 n)}{n}$, that matches the contribution of defects realised as solutions to 5d minimal gauged supergravity found in \cite{Conti:2025wwf}.


\begin{thebibliography}{99}

\bibitem{DHoker:2007zhm}
E.~D'Hoker, J.~Estes and M.~Gutperle,
``Exact half-BPS Type IIB interface solutions. I. Local solution and supersymmetric Janus,''
JHEP \textbf{06} (2007), 021
[arXiv:0705.0022 [hep-th]].

\bibitem{DHoker:2007hhe}
E.~D'Hoker, J.~Estes and M.~Gutperle,
``Exact half-BPS Type IIB interface solutions. II. Flux solutions and multi-Janus,''
JHEP \textbf{06} (2007), 022
[arXiv:0705.0024 [hep-th]].

\bibitem{DHoker:2007mci}
E.~D'Hoker, J.~Estes and M.~Gutperle,
``Gravity duals of half-BPS Wilson loops,''
JHEP \textbf{06} (2007), 063
[arXiv:0705.1004 [hep-th]].

\bibitem{DHoker:2008lup}
E.~D'Hoker, J.~Estes, M.~Gutperle and D.~Krym,
``Exact Half-BPS Flux Solutions in M-theory. I: Local Solutions,''
JHEP \textbf{08} (2008), 028
[arXiv:0806.0605 [hep-th]].

\bibitem{DHoker:2008rje}
E.~D'Hoker, J.~Estes, M.~Gutperle and D.~Krym,
``Exact Half-BPS Flux Solutions in M-theory II: Global solutions asymptotic to AdS(7) x S**4,''
JHEP \textbf{12} (2008), 044
[arXiv:0810.4647 [hep-th]].

\bibitem{Dibitetto:2017tve}
G.~Dibitetto and N.~Petri,
``BPS objects in D = 7 supergravity and their M-theory origin,''
JHEP \textbf{12} (2017), 041
[arXiv:1707.06152 [hep-th]].

\bibitem{Dibitetto:2017klx}
G.~Dibitetto and N.~Petri,
``6d surface defects from massive type IIA,''
JHEP \textbf{01} (2018), 039
[arXiv:1707.06154 [hep-th]].

\bibitem{Dibitetto:2018iar}
G.~Dibitetto and N.~Petri,
``Surface defects in the D4 $-$ D8 brane system,''
JHEP \textbf{01} (2019), 193
[arXiv:1807.07768 [hep-th]].

\bibitem{Dibitetto:2018gtk}
G.~Dibitetto and N.~Petri,
``AdS$_{2}$ solutions and their massive IIA origin,''
JHEP \textbf{05} (2019), 107
[arXiv:1811.11572 [hep-th]].

\bibitem{Gutperle:2018fea}
M.~Gutperle and M.~Vicino,
``Conformal defect solutions in $N=2,D=4$ gauged supergravity,''
Nucl. Phys. B \textbf{942} (2019), 149-163
[arXiv:1811.04166 [hep-th]].

\bibitem{Gutperle:2019dqf}
M.~Gutperle and M.~Vicino,
``Holographic Surface Defects in $D=5$, $N=4$ Gauged Supergravity,''
Phys. Rev. D \textbf{101} (2020) no.6, 066016
[arXiv:1911.02185 [hep-th]].

\bibitem{Chen:2019qib}
K.~Chen and M.~Gutperle,
``Holographic line defects in F(4) gauged supergravity,''
Phys. Rev. D \textbf{100} (2019) no.12, 126015
[arXiv:1909.11127 [hep-th]].

\bibitem{Chen:2020mtv}
K.~Chen, M.~Gutperle and M.~Vicino,
``Holographic Line Defects in $D=4$, $N=2$ Gauged Supergravity,''
Phys. Rev. D \textbf{102} (2020) no.2, 026025
[arXiv:2005.03046 [hep-th]].

\bibitem{Faedo:2020nol}
F.~Faedo, Y.~Lozano and N.~Petri,
``Searching for surface defect CFTs within AdS$_3$,''
JHEP \textbf{11} (2020), 052
[arXiv:2007.16167 [hep-th]].

\bibitem{Lozano:2020sae}
Y.~Lozano, C.~Nunez, A.~Ramirez and S.~Speziali,
``AdS$_{2}$ duals to ADHM quivers with Wilson lines,''
JHEP \textbf{03} (2021), 145
[arXiv:2011.13932 [hep-th]].

\bibitem{Faedo:2020lyw}
F.~Faedo, Y.~Lozano and N.~Petri,
``New $\mathcal{N}=(0,4)$ AdS$_3$ near-horizons in Type IIB,''
JHEP \textbf{04} (2021), 028
[arXiv:2012.07148 [hep-th]].

\bibitem{Gutperle:2020rty}
M.~Gutperle and C.~F.~Uhlemann,
``Surface defects in holographic 5d SCFTs,''
JHEP \textbf{04} (2021), 134
[arXiv:2012.14547 [hep-th]].

\bibitem{Lozano:2021fkk}
Y.~Lozano, N.~Petri and C.~Risco,
``New AdS$_{2}$ supergravity duals of 4d SCFTs with defects,''
JHEP \textbf{10} (2021), 217
[arXiv:2107.12277 [hep-th]].

\bibitem{Gutperle:2022pgw}
M.~Gutperle and N.~Klein,
``A note on co-dimension 2 defects in N=4, d=7 gauged supergravity,''
Nucl. Phys. B \textbf{984} (2022), 115969
[arXiv:2203.13839 [hep-th]].

\bibitem{Lozano:2022vsv}
Y.~Lozano, N.~Petri and C.~Risco,
``Line defects as brane boxes in Gaiotto-Maldacena geometries,''
JHEP \textbf{02} (2023), 193
[arXiv:2212.10398 [hep-th]].

\bibitem{Lozano:2022swp}
Y.~Lozano, N.~Petri and C.~Risco,
``AdS2 near-horizons, defects, and string dualities,''
Phys. Rev. D \textbf{107} (2023) no.10, 106012
[arXiv:2212.11095 [hep-th]].

\bibitem{Conti:2023naw}
A.~Conti, Y.~Lozano and N.~T.~Macpherson,
``New AdS$_{2}$/CFT$_{1}$ pairs from AdS$_{3}$ and monopole bubbling,''
JHEP \textbf{07} (2023), 041
[arXiv:2304.11003 [hep-th]].

\bibitem{Gutperle:2023yrd}
M.~Gutperle, N.~Klein and D.~Rathore,
``Holographic 6d co-dimension 2 defect solutions in M-theory,''
JHEP \textbf{11} (2023), 191
[arXiv:2304.12899 [hep-th]].

\bibitem{Capuozzo:2024onf}
P.~Capuozzo, J.~Estes, B.~Robinson and B.~Suzzoni,
``From large to small $ \mathcal{N} $ = (4, 4) superconformal surface defects in holographic 6d SCFTs,''
JHEP \textbf{08} (2024), 094
[arXiv:2402.11745 [hep-th]].

\bibitem{Lozano:2024idt}
Y.~Lozano, N.~T.~Macpherson, N.~Petri and A.~Ram{\'\i}rez,
``Holographic $ \frac{1}{2} $-BPS surface defects in ABJM,''
JHEP \textbf{08} (2024), 044
[arXiv:2404.17469 [hep-th]].

\bibitem{Arav:2024exg}
I.~Arav, J.~P.~Gauntlett, Y.~Jiao, M.~M.~Roberts and C.~Rosen,
``Superconformal monodromy defects in $ \mathcal{N} $=4 SYM and LS theory,''
JHEP \textbf{08} (2024), 177
[arXiv:2405.06014 [hep-th]].

\bibitem{Karndumri:2024jib}
P.~Karndumri,
``Janus and RG-flow interfaces from matter-coupled F(4) gauged supergravity,''
Phys. Rev. D \textbf{111} (2025) no.2, 026013
[arXiv:2405.17169 [hep-th]].

\bibitem{Karndumri:2024gtv}
P.~Karndumri,
``Line and surface defects in 5D $N=2$ SCFT from matter-coupled F(4) gauged supergravity,''
Eur. Phys. J. C \textbf{84} (2024) no.12, 1268
[arXiv:2406.18946 [hep-th]].

\bibitem{Lozano:2022ouq}
Y.~Lozano, N.~T.~Macpherson, N.~Petri and C.~Risco,
``New AdS$_{3}$/CFT$_{2}$ pairs in massive IIA with (0, 4) and (4, 4) supersymmetries,''
JHEP \textbf{09} (2022), 130
[arXiv:2206.13541 [hep-th]].

\bibitem{Conti:2024rwd}
A.~Conti, G.~Dibitetto, Y.~Lozano, N.~Petri and A.~Ram{\'\i}rez,
``Half-BPS Janus solutions in AdS$_{7}$,''
JHEP \textbf{12} (2024), 198
[arXiv:2407.21619 [hep-th]].

\bibitem{Arav:2024wyg}
I.~Arav, J.~P.~Gauntlett, Y.~Jiao, M.~M.~Roberts and C.~Rosen,
``Superconformal monodromy defects in ABJM and mABJM theory,''
JHEP \textbf{11} (2024), 008
[arXiv:2408.11088 [hep-th]].

\bibitem{Bomans:2024vii}
P.~Bomans and L.~Tranchedone,
``Holographic generalised Gukov-Witten defects,''
JHEP \textbf{03} (2025), 118
[arXiv:2410.18172 [hep-th]].

\bibitem{Faedo:2025kjf}
F.~Faedo, N.~Petri and A.~Segati,
``Defects in 4d SCFTs from supergravity and holographic renormalization,''
JHEP \textbf{05} (2025), 070
[arXiv:2501.17923 [hep-th]].

\bibitem{Conti:2024qgx}
A.~Conti, G.~Dibitetto, Y.~Lozano, N.~Petri and A.~Ram{\'\i}rez,
JHEP \textbf{11} (2024), 131
doi:10.1007/JHEP11(2024)131
[arXiv:2407.21627 [hep-th]].

\bibitem{Ferrero:2020laf}
P.~Ferrero, J.~P.~Gauntlett, J.~M.~P{\'e}rez Ipi{\~n}a, D.~Martelli and J.~Sparks,
``D3-Branes Wrapped on a Spindle,''
Phys. Rev. Lett. \textbf{126} (2021) no.11, 111601
[arXiv:2011.10579 [hep-th]].

\bibitem{Ferrero:2020twa}
P.~Ferrero, J.~P.~Gauntlett, J.~M.~P.~Ipi{\~n}a, D.~Martelli and J.~Sparks,
``Accelerating black holes and spinning spindles,''
Phys. Rev. D \textbf{104} (2021) no.4, 046007
[arXiv:2012.08530 [hep-th]].

\bibitem{Boido:2021szx}
A.~Boido, J.~M.~P.~Ipi{\~n}a and J.~Sparks,
``Twisted D3-brane and M5-brane compactifications from multi-charge spindles,''
JHEP \textbf{07} (2021), 222
[arXiv:2104.13287 [hep-th]].

\bibitem{Ferrero:2021wvk}
P.~Ferrero, J.~P.~Gauntlett, D.~Martelli and J.~Sparks,
``M5-branes wrapped on a spindle,''
JHEP \textbf{11} (2021), 002
[arXiv:2105.13344 [hep-th]].

\bibitem{Ferrero:2021ovq}
P.~Ferrero, M.~Inglese, D.~Martelli and J.~Sparks,
``Multicharge accelerating black holes and spinning spindles,''
Phys. Rev. D \textbf{105} (2022) no.12, 126001
[arXiv:2109.14625 [hep-th]].

\bibitem{Couzens:2021rlk}
C.~Couzens, K.~Stemerdink and D.~van de Heisteeg,
``M2-branes on discs and multi-charged spindles,''
JHEP \textbf{04} (2022), 107
[arXiv:2110.00571 [hep-th]].

\bibitem{Faedo:2021nub}
F.~Faedo and D.~Martelli,
``D4-branes wrapped on a spindle,''
JHEP \textbf{02} (2022), 101
[arXiv:2111.13660 [hep-th]].

\bibitem{Ferrero:2021etw}
P.~Ferrero, J.~P.~Gauntlett and J.~Sparks,
``Supersymmetric spindles,''
JHEP \textbf{01} (2022), 102
[arXiv:2112.01543 [hep-th]].

\bibitem{Bianchi:2019sxz}
L.~Bianchi and M.~Lemos,
``Superconformal surfaces in four dimensions,''
JHEP \textbf{06} (2020), 056
[arXiv:1911.05082 [hep-th]].

\bibitem{Bianchi:2021snj}
L.~Bianchi, A.~Chalabi, V.~Proch{\'a}zka, B.~Robinson and J.~Sisti,
``Monodromy defects in free field theories,''
JHEP \textbf{08} (2021), 013
[arXiv:2104.01220 [hep-th]].

\bibitem{Giombi:2021uae}
S.~Giombi, E.~Helfenberger, Z.~Ji and H.~Khanchandani,
``Monodromy defects from hyperbolic space,''
JHEP \textbf{02} (2022), 041
[arXiv:2102.11815 [hep-th]].

\bibitem{ContiStuardo}
A.~Conti, R.~Stuardo,
``Monodromy Defects in Maximally Supersymmetric Yang-Mills Theories from Holography,''
To appear

\bibitem{Apruzzi:2013yva}
F.~Apruzzi, M.~Fazzi, D.~Rosa and A.~Tomasiello,
``All AdS$_7$ solutions of type II supergravity,''
JHEP \textbf{04} (2014), 064
[arXiv:1309.2949 [hep-th]].

\bibitem{Brandhuber:1999np}
A.~Brandhuber and Y.~Oz,
``The D-4 - D-8 brane system and five-dimensional fixed points,''
Phys. Lett. B \textbf{460} (1999), 307-312
[arXiv:hep-th/9905148 [hep-th]].

\bibitem{Ryu:2006bv}
S.~Ryu and T.~Takayanagi,
Phys. Rev. Lett. \textbf{96} (2006), 181602
doi:10.1103/PhysRevLett.96.181602
[arXiv:hep-th/0603001 [hep-th]].

\bibitem{Jensen:2018rxu}
K.~Jensen, A.~O'Bannon, B.~Robinson and R.~Rodgers,
``From the Weyl Anomaly to Entropy of Two-Dimensional Boundaries and Defects,''
Phys. Rev. Lett. \textbf{122} (2019) no.24, 241602
[arXiv:1812.08745 [hep-th]].

\bibitem{Kobayashi:2018lil}
N.~Kobayashi, T.~Nishioka, Y.~Sato and K.~Watanabe,
``Towards a $C$-theorem in defect CFT,''
JHEP \textbf{01} (2019), 039
[arXiv:1810.06995 [hep-th]].

\bibitem{Chalabi:2021jud}
A.~Chalabi, C.~P.~Herzog, A.~O'Bannon, B.~Robinson and J.~Sisti,
``Weyl anomalies of four dimensional conformal boundaries and defects,''
JHEP \textbf{02} (2022), 166
[arXiv:2111.14713 [hep-th]].

\bibitem{Conti:2025wwf}
A.~Conti, Y.~Lozano, F.~Rogdakis and C.~Rosen,
``Defect entanglement entropy for superconformal monodromy defects,''
[arXiv:2511.22695 [hep-th]].

\bibitem{Apruzzi:2015zna}
F.~Apruzzi, M.~Fazzi, A.~Passias and A.~Tomasiello,
``Supersymmetric AdS$_{5}$ solutions of massive IIA supergravity,''
JHEP \textbf{06} (2015), 195
[arXiv:1502.06620 [hep-th]].

\bibitem{Jensen:2013lxa}
K.~Jensen and A.~O'Bannon,
``Holography, Entanglement Entropy, and Conformal Field Theories with Boundaries or Defects,''
Phys. Rev. D \textbf{88} (2013) no.10, 106006
[arXiv:1309.4523 [hep-th]].

\bibitem{Affleck:1991tk}
I.~Affleck and A.~W.~W.~Ludwig,
``Universal noninteger 'ground state degeneracy' in critical quantum systems,''
Phys. Rev. Lett. \textbf{67} (1991), 161-164

\bibitem{Brunner:1997gf}
I.~Brunner and A.~Karch,
``Branes at orbifolds versus Hanany Witten in six-dimensions,''
JHEP \textbf{03} (1998), 003
[arXiv:hep-th/9712143 [hep-th]].

\bibitem{Hanany:1997gh}
A.~Hanany and A.~Zaffaroni,
``Branes and six-dimensional supersymmetric theories,''
Nucl. Phys. B \textbf{529} (1998), 180-206
[arXiv:hep-th/9712145 [hep-th]].

\bibitem{Gaiotto:2014lca}
D.~Gaiotto and A.~Tomasiello,
``Holography for (1,0) theories in six dimensions,''
JHEP \textbf{12} (2014), 003
[arXiv:1404.0711 [hep-th]].

\bibitem{Cremonesi:2015bld}
S.~Cremonesi and A.~Tomasiello,
``6d holographic anomaly match as a continuum limit,''
JHEP \textbf{05} (2016), 031
[arXiv:1512.02225 [hep-th]].

\bibitem{Cvetic:1999xp}
M.~Cvetic, M.~J.~Duff, P.~Hoxha, J.~T.~Liu, H.~Lu, J.~X.~Lu, R.~Martinez-Acosta, C.~N.~Pope, H.~Sati and T.~A.~Tran,
``Embedding AdS black holes in ten-dimensions and eleven-dimensions,''
Nucl. Phys. B \textbf{558} (1999), 96-126
[arXiv:hep-th/9903214 [hep-th]].

\bibitem{Lin:2004nb}
H.~Lin, O.~Lunin and J.~M.~Maldacena,
``Bubbling AdS space and 1/2 BPS geometries,''
JHEP \textbf{10} (2004), 025
[arXiv:hep-th/0409174 [hep-th]].

\bibitem{Lu:2003iv}
H.~Lu, C.~N.~Pope and J.~F.~Vazquez-Poritz,
``From AdS black holes to supersymmetric flux branes,''
Nucl. Phys. B \textbf{709} (2005), 47-68
[arXiv:hep-th/0307001 [hep-th]].

\bibitem{Liu:1999ai}
J.~T.~Liu and R.~Minasian,
``Black holes and membranes in AdS(7),''
Phys. Lett. B \textbf{457} (1999), 39-46
[arXiv:hep-th/9903269 [hep-th]].

\bibitem{Passias:2015gya}
A.~Passias, A.~Rota and A.~Tomasiello,
``Universal consistent truncation for 6d/7d gauge/gravity duals,''
JHEP \textbf{10} (2015), 187
[arXiv:1506.05462 [hep-th]].

\bibitem{Nunez:2018ags}
C.~N{\'u}{\~n}ez, J.~M.~Pen{\'\i}n, D.~Roychowdhury and J.~Van Gorsel,
``The non-Integrability of Strings in Massive Type IIA and their Holographic duals,''
JHEP \textbf{06} (2018), 078
[arXiv:1802.04269 [hep-th]].

\bibitem{Bobev:2016phc}
N.~Bobev, G.~Dibitetto, F.~F.~Gautason and B.~Truijen,
``Holography, Brane Intersections and Six-dimensional SCFTs,''
JHEP \textbf{02} (2017), 116
[arXiv:1612.06324 [hep-th]].

\bibitem{Brandhuber:1998xy}
A.~Brandhuber, N.~Itzhaki, J.~Sonnenschein and S.~Yankielowicz,
``Baryons from supergravity,''
JHEP \textbf{07} (1998), 020
[arXiv:hep-th/9806158 [hep-th]].

\bibitem{Maldacena:1998im}
J.~M.~Maldacena,
``Wilson loops in large N field theories,''
Phys. Rev. Lett. \textbf{80} (1998), 4859-4862
[arXiv:hep-th/9803002 [hep-th]].

\bibitem{Merrikin:2022yho}
P.~Merrikin, C.~Nunez and R.~Stuardo,
``Compactification of 6d N=(1,0) quivers, 4d SCFTs and their holographic dual Massive IIA backgrounds,''
Nucl. Phys. B \textbf{996} (2023), 116356
[arXiv:2210.02458 [hep-th]].

\bibitem{Cheung:2022ilc}
K.~C.~M.~Cheung, J.~H.~T.~Fry, J.~P.~Gauntlett and J.~Sparks,
``M5-branes wrapped on four-dimensional orbifolds,''
JHEP \textbf{08} (2022), 082
[arXiv:2204.02990 [hep-th]].

\bibitem{Suh:2022olh}
M.~Suh,
``M5-branes and D4-branes wrapped on a direct product of spindle and Riemann surface,''
JHEP \textbf{02} (2024), 205
[arXiv:2207.00034 [hep-th]].

\bibitem{Faedo:2019cvr}
A.~F.~Faedo, C.~Nunez and C.~Rosen,
``Consistent truncations of supergravity and $\frac{1}{2}$-BPS RG flows in $4d$ SCFTs,''
JHEP \textbf{03} (2020), 080
[arXiv:1912.13516 [hep-th]].

\bibitem{Seiberg:1996bd}
N.~Seiberg,
``Five-dimensional SUSY field theories, nontrivial fixed points and string dynamics,''
Phys. Lett. B \textbf{388} (1996), 753-760
[arXiv:hep-th/9608111 [hep-th]].

\bibitem{Giri:2021xta}
S.~Giri,
``Black holes with spindles at the horizon,''
JHEP \textbf{06} (2022), 145
[arXiv:2112.04431 [hep-th]].

\bibitem{Hristov:2024qiy}
K.~Hristov and M.~Suh,
``Spindle black holes and theories of class $ \mathcal{F} $,''
JHEP \textbf{08} (2024), 006
[arXiv:2405.17432 [hep-th]].

\bibitem{Couzens:2022lvg}
C.~Couzens, H.~Kim, N.~Kim, Y.~Lee and M.~Suh,
``D4-branes wrapped on four-dimensional orbifolds through consistent truncation,''
JHEP \textbf{02} (2023), 025
[arXiv:2210.15695 [hep-th]].

\bibitem{Faedo:2022rqx}
F.~Faedo, A.~Fontanarossa and D.~Martelli,
``Branes wrapped on orbifolds and their gravitational blocks,''
Lett. Math. Phys. \textbf{113} (2023) no.3, 51
[arXiv:2210.16128 [hep-th]].

\bibitem{DAuria:2000afl}
R.~D'Auria, S.~Ferrara and S.~Vaula,
``Matter coupled F(4) supergravity and the AdS(6) / CFT(5) correspondence,''
JHEP \textbf{10} (2000), 013
[arXiv:hep-th/0006107 [hep-th]].

\bibitem{Cvetic:1999xx}
M.~Cvetic, S.~S.~Gubser, H.~Lu and C.~N.~Pope,
``Symmetric potentials of gauged supergravities in diverse dimensions and Coulomb branch of gauge theories,''
Phys. Rev. D \textbf{62} (2000), 086003
[arXiv:hep-th/9909121 [hep-th]].

\bibitem{Cvetic:1999un}
M.~Cvetic, H.~Lu and C.~N.~Pope,
``Gauged six-dimensional supergravity from massive type IIA,''
Phys. Rev. Lett. \textbf{83} (1999), 5226-5229
[arXiv:hep-th/9906221 [hep-th]].

\bibitem{Jafferis:2012iv}
D.~L.~Jafferis and S.~S.~Pufu,
``Exact results for five-dimensional superconformal field theories with gravity duals,''
JHEP \textbf{05} (2014), 032
[arXiv:1207.4359 [hep-th]].

\bibitem{Capuozzo:2023fll}
P.~Capuozzo, J.~Estes, B.~Robinson and B.~Suzzoni,
``Holographic Weyl anomalies for 4d defects in 6d SCFTs,''
JHEP \textbf{04} (2024), 120
[arXiv:2310.17447 [hep-th]].

\bibitem{Bah:2018lyv}
I.~Bah, A.~Passias and P.~Weck,
``Holographic duals of five-dimensional SCFTs on a Riemann surface,''
JHEP \textbf{01} (2019), 058
[arXiv:1807.06031 [hep-th]].

\bibitem{Hosseini:2018usu}
S.~M.~Hosseini, K.~Hristov, A.~Passias and A.~Zaffaroni,
``6D attractors and black hole microstates,''
JHEP \textbf{12} (2018), 001
[arXiv:1809.10685 [hep-th]].

\bibitem{Hosseini:2020wag}
S.~M.~Hosseini and K.~Hristov,
``4d F(4) gauged supergravity and black holes of class $\mathcal{F}$,''
JHEP \textbf{02} (2021), 177
[arXiv:2011.01943 [hep-th]].

\bibitem{Karndumri:2015eta}
P.~Karndumri,
``Twisted compactification of N = 2 5D SCFTs to three and two dimensions from F(4) gauged supergravity,''
JHEP \textbf{09} (2015), 034
[arXiv:1507.01515 [hep-th]].

\bibitem{Cvetic:1999ne}
M.~Cvetic and S.~S.~Gubser,
``Phases of R charged black holes, spinning branes and strongly coupled gauge theories,''
JHEP \textbf{04} (1999), 024
[arXiv:hep-th/9902195 [hep-th]].

\bibitem{Herzog:2019rke}
C.~P.~Herzog and I.~Shamir,
``How a-type anomalies can depend on marginal couplings,''
Phys. Rev. Lett. \textbf{124} (2020) no.1, 011601
[arXiv:1907.04952 [hep-th]].

\bibitem{Santilli:2023fuh}
L.~Santilli and C.~F.~Uhlemann,
``3d defects in 5d: RG flows and defect F-maximization,''
JHEP \textbf{06} (2023), 136
[arXiv:2305.01004 [hep-th]].

\bibitem{Bah:2012dg}
I.~Bah, C.~Beem, N.~Bobev and B.~Wecht,
``Four-Dimensional SCFTs from M5-Branes,''
JHEP \textbf{06} (2012), 005
[arXiv:1203.0303 [hep-th]].


\end{thebibliography}
\end{document}